\documentclass[twocolumn]{aastex701}

\usepackage{amsmath}
\usepackage{amssymb}
\usepackage{placeins}

\DeclareMathOperator{\sech}{sech}

\begin{document}

\title{Tearing Instability in Gyrotropic MHD: Effects of Equilibrium Pressure Anisotropy}

\author[orcid=0000-0002-0176-9909,sname='Kowal']{Grzegorz Kowal}
\affiliation{Escola de Artes, Ciências e Humanidades, University of São Paulo, São Paulo, SP, Brazil}
\email{grzegorz.kowal@usp.br}

\author[orcid=0000-0001-7144-5777,sname='Ferreira-Santos']{Gabriel L. Ferreira-Santos}
\affiliation{Astrophysics Division, National Institute for Space Research, São José dos Campos, SP, Brazil}
\email{gabriel.ferreira@inpe.br}

\author[orcid=0000-0002-1914-6654,sname='Falceta-Gonçalves']{Diego A. Falceta-Gonçalves}
\affiliation{Escola de Artes, Ciências e Humanidades, University of São Paulo, São Paulo, SP, Brazil}
\email{dfalcetal@usp.br}

\correspondingauthor{Grzegorz Kowal}

\begin{abstract}
Weakly collisional plasmas are widespread in astrophysics and can sustain pressure anisotropy, yet most analytical tearing-mode scalings assume an isotropic equilibrium. We develop a linear theory of resistive tearing in nonideal gyrotropic MHD for a force-free Harris current sheet characterized by perpendicular plasma beta $\beta_0$ and parallel-minus-perpendicular beta difference $\Delta\beta_0$. In the ideal outer region, anisotropy changes the far-field decay rate, the matching parameter $\Delta'$, and sets the upper wavenumber cutoff for localized tearing instability, $\alpha\equiv ka<\alpha_c=\sqrt{\mathcal{A}/\mathcal{R}_0}$, where $\mathcal{A}=1-\Delta\beta_0/2$ and $\mathcal{R}_0=1+\frac{1}{2}[(\gamma_\parallel+\gamma_\perp-2)\beta_0+\gamma_\parallel\Delta\beta_0]$, and $\gamma_\parallel$ and $\gamma_\perp$ are the parallel and perpendicular polytropic indices specifying the gyrotropic pressure closure. In the resistive inner layer, anisotropy enters the leading momentum balance through $\mathcal{A}$. We derive modified FKR and Coppi branches. By matching the branches at their crossover wavenumber, we identify the fastest-growing mode and find the maximum-growth-rate scaling $\gamma_{\max}\tau_A\sim\mathcal{A}^{1/2}\mathcal{R}_0^{-1/4}S^{-1/2}$, where $S$ is the Lundquist number and $\tau_A$ is the Alfv\'en crossing time. Thus the classical Lundquist-number exponent is retained, while the prefactor depends on the equilibrium anisotropy, plasma-$\beta$, and gyrotropic closure. PSECAS eigenvalue calculations support the Coppi branch and are consistent with the FKR branch when a fitted finite-wavelength approximation for $\Delta'$ is used. Within the localized-mode and pressure-positive domain, positive $\Delta\beta_0$ generally suppresses tearing and broadens the inner layer, whereas negative $\Delta\beta_0$ enhances growth and shifts the fastest mode to larger wavenumber. By identifying how prescribed equilibrium pressure anisotropy modifies both ideal outer matching and resistive inner-layer dynamics, this work provides a controlled extension of classical tearing theory within the gyrotropic-MHD regime.
\end{abstract}


\section{Introduction}
\label{sec:introduction}

The tearing instability is one of the fundamental mechanisms by which thin
current sheets become unstable and initiate magnetic reconnection. In
classical resistive MHD theory, \citet{1963PhFl....6..459F} showed that finite
magnetic diffusivity destabilizes planar current sheets, leading to the growth
of magnetic islands. The corresponding boundary-layer theory separates the
system into an ideal outer region, where the tearing stability index
$\Delta'$ is determined, and a resistive inner layer, where reconnection takes
place. This analysis yields the familiar constant-$\psi$ scaling of the FKR
branch, the nonconstant-$\psi$ scaling associated with the Coppi regime
\citep{1976SvJPP...2..533C}, and the maximum-growth estimates
$\gamma_{\max}\tau_A\sim S^{-1/2}$ and $k_{\max}a\sim S^{-1/4}$. Here
$S$ is the Lundquist number and $\tau_A$ is the Alfv\'en crossing time.
Within incompressible resistive MHD, these leading scalings are independent of
the plasma-$\beta$, defined as the ratio of thermal pressure to magnetic
pressure, $\beta=2\mu_0p/B^2$.

This plasma-$\beta$ independence is a direct consequence of the isotropic
pressure closure used in standard MHD. Many space and astrophysical plasmas,
however, are weakly collisional, so collisions may be insufficient to maintain
an isotropic pressure tensor. In such magnetized systems, the thermal pressure
is more naturally decomposed into components parallel and perpendicular to the
local magnetic field. The double-adiabatic CGL model
\citep{1956RSPSA.236..112C} provides the simplest fluid description of this
gyrotropic response, while extended formulations including dissipative heating
restore the energy balance required in nonideal systems
\citep{2002PhPl....9.2455H}. These considerations motivate the use of a
nonideal gyrotropic MHD framework to study how pressure anisotropy modifies the
large-scale linear onset of tearing, before kinetic effects become dominant
inside thinner diffusion regions.

The role of pressure anisotropy in tearing has been investigated from both
kinetic and fluid perspectives. Collisionless studies of anisotropic
Harris-type neutral sheets showed that temperature anisotropy can change the
mode structure, growth rate, and preferred wavelength, and that particle-orbit
physics outside the electron diffusion region can be important for the tearing
eigenmode \citep{ChenPalmadesso1984,Quest2010}. In parallel,
gyrotropic-MHD studies found that the tearing response depends sensitively on
the closure used for $p_\parallel$ and $p_\perp$, on the plasma-$\beta$, and on
the magnetic geometry; in some regimes, the resulting modes become oscillatory
or grow on Alfv\'enic time scales \citep{ChiouHau2002,ChiouHau2003}. These
studies establish the importance of anisotropy, but they do not provide the
outer--inner asymptotic construction needed to determine separately how a
prescribed equilibrium anisotropy modifies the ideal matching parameter
$\Delta'$ and the resistive inner-layer force balance. Consequently, its
effects on the FKR, Coppi, and fastest-growing-mode scalings remain to be
established. Addressing this gap is the central objective of the present work.

This question is also connected to the onset of reconnection in space and
astrophysical plasmas. Classical tearing scalings are widely used to interpret
the disruption of thin current sheets and the transition to fast reconnection
in solar and magnetospheric environments
\citep{PucciVelli2014,LoureiroUzdensky2016,UzdenskyLoureiro2016}. Hybrid and
kinetic studies further show that pressure anisotropy can either enhance or
delay current-sheet tearing in the solar wind \citep{Matteini2013}, and that
high-plasma-$\beta$ pressure anisotropy and mirror fluctuations can reshape a
forming current sheet and trigger tearing earlier than expected from an
isotropic Harris profile \citep{AltKunz2019,WinartoKunz2022}. These results
suggest that pressure anisotropy is not only a kinetic-scale consequence of
reconnection, but can also influence the large-scale stability problem that
sets the tearing onset.

Motivated by this unresolved theoretical gap and its relevance to reconnection
onset, we extend the isotropic-equilibrium analysis of
\citet{2025ApJ...993...74F} to a force-free Harris-type current sheet with
prescribed equilibrium pressure anisotropy. Its uniform magnetic-field
magnitude and uniform parallel and perpendicular pressures isolate the
anisotropic correction from equilibrium pressure-gradient forces, allowing us
to determine its separate effects on ideal outer matching and resistive
inner-layer dynamics. This controlled model is intended to clarify the linear
resistive gyrotropic-MHD regime rather than provide a direct model of
astrophysical reconnection; it excludes kinetic anisotropy relaxation, Hall
physics, compressibility, and nonlinear island evolution.

The paper is organized as follows. Section~\ref{sec:equations} presents the
nonideal gyrotropic MHD equations and the linearized perturbation system for
the anisotropic equilibrium. Section~\ref{sec:theory} derives the ideal
outer-region equation, the tearing stability index, the leading inner-layer
equations, the FKR and Coppi branches, and the fastest-growing-mode scalings.
Section~\ref{sec:numerical-results} describes the numerical eigenvalue
calculation and compares the numerical dispersion relations and parameter
scans with the analytical predictions.
Section~\ref{sec:discussion} discusses the implications and
limitations of the results, and Section~\ref{sec:conclusions} summarizes the
main conclusions. The appendices collect the equilibrium identities
(Appendix~\ref{app:equilibrium-identities}), the derivation of the linearized
system including its real-amplitude representation
(Appendix~\ref{app:linearization}), the double-isothermal outer problem
(Appendix~\ref{app:isothermal-outer}), and the numerical check of the
outer-region stability boundary
(Appendix~\ref{app:outer-stability-boundary}).

\section{Nonideal Gyrotropic MHD Model and Linearized Equations}
\label{sec:equations}

Following \cite{2025ApJ...993...74F}, we use the incompressible
($\nabla\cdot\mathbf{v}=0$) nonideal gyrotropic MHD model. The equations are
written in Alfv\'enic units: the magnetic field is normalized by the
upstream magnetic-field strength $B_{\rm up}$, the density by the uniform
upstream density $\rho_0$, the velocity by
$v_A=B_{\rm up}/\sqrt{\mu_0\rho_0}$, and the pressures by
$B_{\rm up}^2/\mu_0$. Thus $\mu_0=\rho_0=1$ in the equations below, while the
current-sheet thickness $a$ remains explicit. Lengths and times can
subsequently be normalized by $a$ and $\tau_A=a/v_A$, respectively. The
governing momentum and induction equations are
\begin{equation}
\frac{\partial \mathbf{v}}{\partial t} +(\mathbf{v} \cdot \nabla ) \mathbf{v} = - \nabla \cdot \left( p_\perp \mathbf{I} + \Delta p \mathbf{b} \mathbf{b} \right) + \mathbf{J} \times \mathbf{B} + \nu \nabla^2 \mathbf{v}
\label{eq:momentum}
\end{equation}
\begin{equation}
\frac{\partial \mathbf{B}}{\partial t} =\nabla\times(\mathbf{v}\times\mathbf{B}) +\eta \nabla ^{2}\mathbf{B}, \quad \nabla \cdot \mathbf{B} = 0
\label{eq:induction}
\end{equation}
Here $\mathbf{J}=\nabla\times\mathbf{B}$,
$\Delta p=p_\parallel-p_\perp$, and
$\mathbf{b}=\mathbf{B}/|\mathbf{B}|$. The coefficients $\nu$ and $\eta$ are
the kinematic viscosity and magnetic diffusivity, respectively. Their
dimensionless values in current-sheet units are
$\eta/(av_A)=S^{-1}$ and
$\nu/(av_A)=Pr_mS^{-1}$, where
$S=av_A/\eta$ and $Pr_m=\nu/\eta$. In the numerical calculations, we set
$a=v_A=1$, so the dimensionless transport coefficients are written simply as
$\eta=S^{-1}$ and $\nu=Pr_mS^{-1}$.

The nonideal gyrotropic MHD equations are closed by incorporating the evolution
equations for the two pressure tensor components, as described
in \cite{2002PhPl....9.2455H}:
\begin{align}
\frac{\partial p_\parallel}{\partial t}
+(\mathbf{v}\cdot\nabla)p_\parallel
&=-(\gamma_\parallel-1)p_\parallel
\left[\mathbf{b}\cdot(\mathbf{b}\cdot\nabla\mathbf{v})\right] \nonumber\\
&+\eta(\gamma_\parallel-1)(\mathbf{b}\cdot\mathbf{J})^2
\nonumber\\
&+\frac{1}{3}\nu(\gamma_\parallel-1)(\nabla\times\mathbf{v})^2 ,
\label{eq:parallel_pressure}\\
\frac{\partial p_\perp}{\partial t}
+(\mathbf{v}\cdot\nabla)p_\perp
&=(\gamma_\perp-1)p_\perp
\left[\mathbf{b}\cdot(\mathbf{b}\cdot\nabla\mathbf{v})\right] \nonumber\\
&+\eta(\gamma_\perp-1)
\bigl[\mathbf{J}\cdot\mathbf{J}-(\mathbf{b}\cdot\mathbf{J})^2\bigr]
\nonumber\\
&+\frac{2}{3}\nu(\gamma_\perp-1)(\nabla\times\mathbf{v})^2 ,
\label{eq:perp_pressure}
\end{align}
where $\gamma_\parallel$ and $\gamma_\perp$ are the parallel and perpendicular
polytropic indices specifying the gyrotropic pressure closure.

The double-adiabatic CGL formulation \citep{1956RSPSA.236..112C} describes an
ideal collisionless response and therefore does not account for the conversion
of magnetic and kinetic energy into thermal energy by nonideal dissipation.
Following the energy-law extension of \citet{2002PhPl....9.2455H}, the pressure
equations above include Ohmic and viscous heating. With the corresponding
thermal-energy definition and suitable boundary conditions, these source terms
track the transfer from magnetic and kinetic reservoirs and preserve the
global total-energy balance.

We assume a 2.5D geometry in the $x$--$z$ plane, with all $y$-derivatives
equal to zero. The prescribed current sheet has
$\mathbf{v}_0=0$ and
$\mathbf{B}_0=B_0(z)\hat{\mathbf{i}}+B_g(z)\hat{\mathbf{j}}$, where
$B_0(z)=\tanh(z/a)$ and $B_g(z)=\sech(z/a)$, so that
$|\mathbf{B}_0|^2=1$. The uniform equilibrium pressures are
$p_{\perp,0}=\beta_0/2$ and
$p_{\parallel,0}=(\beta_0+\Delta\beta_0)/2$. Pressure positivity therefore
requires $\beta_0>0$ and $\Delta\beta_0>-\beta_0$.

For finite $\eta$, this prescribed profile is not an exact stationary solution
of the nonideal equations: $\eta\nabla^2\mathbf{B}_0\neq0$, and the parallel
Ohmic-heating source is nonzero when $\gamma_\parallel\neq1$. We therefore
adopt the standard frozen-equilibrium tearing ordering: the zeroth-order
nonideal evolution of the prescribed sheet is assumed to be externally
balanced or negligible on the instability timescale, while the first-order
resistive terms are retained in the perturbation equations. For an unstable
mode, this approximation requires $\gamma\tau_A\gg S^{-1}$ and becomes
nonuniform in the immediate marginal limit $\gamma\to0$.

Small perturbations are taken in the form
$\delta f(t,\mathbf{r})=\delta f(z)\exp(ikx+\sigma t)$, where $k$ is the
$x$-component of the wavevector and $\sigma$ is the complex eigenvalue. The
growth rate is $\gamma\equiv{\rm Re}(\sigma)$, distinct from the closure
indices $\gamma_\parallel$ and $\gamma_\perp$. Linearizing
Eqs.~(\ref{eq:momentum})--(\ref{eq:perp_pressure}) about the frozen profile
yields five coupled equations for the velocity, magnetic-field, and
pressure-anisotropy perturbations:
\begin{widetext}
\begin{align}
\label{dvy}
& \sigma \delta v_y = \underbrace{i k B_0 \delta B_y + B_g' \delta B_z}_{\text{ideal MHD part}} \underbrace{- i k B_0 B_g \delta \Delta p}_{\text{pressure anisotropy fluctuations}} \underbrace{- \frac{\Delta \beta_0}{2} \left[ i k B_0 \left( 1 - 2 B_g^2 \right) \delta B_y + B_g' \delta B_z + 2 B_0^2 B_g \delta B_z' \right]}_{\text{equilibrium pressure anisotropy}} \\
& \hspace{1.2in} \underbrace{+ \nu \left( \delta v_y'' - k^2 \delta v_y \right)}_{\text{viscous part}} \nonumber
\end{align}
\begin{align}
\label{dvz}
& \sigma \left( \delta v_z'' - k^2 \delta v_z \right) =
 \underbrace{i k \left[ B_0 \left( \delta B_z'' - k^2 \delta B_z \right) - B_0'' \delta B_z \right]}_{\text{ideal MHD part}} \underbrace{- k^2 B_0 \left[ 2 B_0' \delta \Delta p + B_0 \delta \Delta p' \right]}_{\text{pressure anisotropy fluctuations}} \\
& \hspace{0.8in} \underbrace{- \frac{\Delta \beta_0}{2} \Bigl\{ i k \left[ B_0 \left( \delta B_z'' - k^2 \delta B_z \right) - B_0'' \delta B_z \right] - 2 \left[ i k B_0^2 \left[ B_0 \delta B_z'' + 3 B_0' \delta B_z' \right] \right.}_{\text{equilibrium pressure anisotropy}} \nonumber \\
& \hspace{0.4in} \quad \underbrace{\left. + k^2 B_0^2 B_g \delta B_y' + k^2 B_0 \left[ B_0 B_g' +2 B_0' B_g \right] \delta B_y \right] \Bigr\}}_{\text{equilibrium pressure anisotropy}} \underbrace{+ \nu \left( \delta v_{z}'''' - 2 k^2 \delta v_{z}'' + k^4 \delta v_{z} \right)}_{\text{viscous part}} \nonumber
\end{align}
\begin{align}
\label{dby}
& \sigma \delta B_y = \underbrace{i k B_0 \delta v_y - B_g' \delta v_z}_{\text{ideal MHD part}} \underbrace{+ \eta \left( \delta B_y'' - k^2 \delta B_y \right)}_{\text{resistive part}}
\end{align}
\begin{align}
\label{dbz}
& \sigma \delta B_z = \underbrace{i k B_0 \delta v_z}_{\text{ideal MHD part}} \underbrace{+ \eta \left( \delta B_z'' - k^2 \delta B_z \right)}_{\text{resistive part}}
\end{align}
\begin{align}
\label{dp_lin}
& \sigma k \delta \Delta p = \underbrace{\frac{1}{2} \left( \gamma_\parallel + \gamma_\perp - 2 \right) \beta_0 k B_0 \left( B_0 \delta v_{z}' - i k B_{g} \delta v_{y} \right)}_{\text{plasma-$\beta$ dependence}} \underbrace{+ \frac{1}{2} \left( \gamma_\parallel - 1 \right) \Delta \beta_0 k B_0 \left( B_0 \delta v_{z}' - i k B_{g} \delta v_{y} \right)}_{\text{equilibrium pressure anisotropy}} \\
& \hspace{0.1in} \underbrace{+ 2 \eta \left( \gamma_\parallel + \gamma_\perp - 2 \right) \left( B_g B_0' - B_0 B_g' \right) \left\{ k \left( B_{0}' \delta B_{y} - B_0 \delta B_{y}' \right) \right.}_{\text{parallel ohmic response}} \nonumber \\
&  \hspace{0.24in} \underbrace{\left. + i \left[ B_g \left( \delta B_{z}'' - k^2 \delta B_{z} \right) - B_{g}' \delta B_{z}' \right] - \left( B_g B_0' - B_0 B_g' \right) \left( i B_0 \delta B_{z}' + k B_g \delta B_{y} \right) \right\}}_{\text{parallel ohmic response}} \nonumber \\
& \hspace{0.1in} \underbrace{- 2 \eta \left( \gamma_\perp - 1 \right) \left[ k B_{g}' \delta B_{y}' + i B_{0}' \left( \delta B_{z}'' - k^{2} \delta B_{z} \right) \right]}_{\text{perpendicular ohmic response}} \nonumber
\end{align}
\end{widetext}
In the above equations, $'$, $''$, and $''''$ denote the first-, second-, and
fourth-order derivatives with respect to $z$, respectively. The details of
this derivation are presented in Appendix~\ref{app:linearization}.

\section{Analytical Theory of the Tearing Instability in Gyrotropic MHD}
\label{sec:theory}

We use the standard outer--inner construction for resistive tearing modes
\citep[e.g.][]{1963PhFl....6..459F,2018JPhCS1100a2003B}, following the
notation of \citet{2025ApJ...993...74F} wherever the analysis is unchanged.
Only the definitions needed for the anisotropic extension are repeated here.
Away from the resonant surface, the outer solution is ideal and determines
the jump condition passed to the resistive layer. For perturbations
proportional to $e^{ikx+\sigma t}$, the reconnecting field perturbation is
related to the usual magnetic flux function by $\delta B_z=ik\psi$. We
therefore define the dimensionless tearing index directly from $\delta B_z$
as
\begin{equation}
  \label{eq:jump}
  \Delta' =
  \frac{a}{\delta B_z(0)}
  \left(\frac{d \delta B_z}{dz}\Big|_{0^+}
- \frac{d \delta B_z}{dz}\Big|_{0^-}\right).
\end{equation}
This normalization makes $\Delta'$ dimensionless, with dimensional jump
$\Delta'/a$. The inner layer then supplies the resistive matching relation:
the constant-$\psi$ ordering gives the FKR branch, the nonconstant-$\psi$
ordering gives the Coppi branch, and their matching identifies the
fastest-growing mode. The details common to the isotropic-equilibrium problem
are not repeated; below we keep only the modifications introduced by a
prescribed equilibrium pressure anisotropy.

\subsection{Outer Region Solution}
\label{ssec:outer}

The derivation of the governing differential equation for the magnetic field
perturbation $\delta B_z$ in the outer ideal region is obtained by ignoring
resistivity and viscosity. Whereas \citet{2025ApJ...993...74F} assumed
$\Delta\beta_0=0$, here we explore arbitrary nonzero $\Delta\beta_0$.
The primary objective is to obtain a single, self-contained second-order
differential equation for $\delta B_z$. In the ideal outer region,
Eq.~(\ref{dp_lin}) reduces to
\begin{align}
    \sigma \delta \Delta p
    &= \bar{\beta} B_0\left(B_0\delta v_z'-ikB_g\delta v_y\right), \\
    \bar{\beta}
    &\equiv \frac{1}{2}\left[
    \left(\gamma_\parallel+\gamma_\perp-2\right)\beta_0
    +\left(\gamma_\parallel-1\right)\Delta\beta_0
    \right].
    \label{eq:barbeta}
\end{align}
The coefficient $\bar{\beta}$ therefore contains the pressure response
associated with both the equilibrium plasma-$\beta$ and the imposed
equilibrium pressure anisotropy.

The analysis begins with the $z$-component of the momentum equation,
Eq.~(\ref{dvz}), and proceeds by expressing the velocity perturbation
$\delta v_z$ in terms of $\delta B_z$ via the ideal version of
Eq.~(\ref{dbz}),
\begin{equation}
    \delta v_z = -\frac{i\sigma}{kB_0}\delta B_z .
\end{equation}
Substitution of this relation and its derivatives into Eq.~(\ref{dvz}) gives
\begin{widetext}
\begin{equation}
\begin{split}
    - & \frac{\sigma^{2}}{k^2 B_0^2} \left[ \delta B_z'' - 2 \frac{B_0'}{B_0} \delta B_z' - \left(\frac{B_0''}{B_0} - 2\frac{B_0'^2}{B_0^2} + k^2 \right) \delta B_z \right] = \\
    & \left( 1 -\frac{1}{2} \Delta \beta_0 +\Delta \beta_0 B_0^2 \right) \delta B_z'' + 3 \Delta \beta_0 B_0 B'_0 \delta B_z' - \left( 1 - \frac{1}{2} \Delta \beta_0 \right) \left( \frac{B_0''}{B_0} + k^2 \right) \delta B_z \\
    & - i k \Delta \beta_0 B_0 B_g \left[ \delta B_y' + \left( \frac{B_g'}{B_g} + 2 \frac{B_0'}{B_0} \right) \delta B_y \right] + i k B_0\left(\delta \Delta p' +2 \frac{B_0'}{B_0}\delta \Delta p \right).
\end{split}
\label{bz-relation}
\end{equation}
\end{widetext}
Because the tearing mode is a resistive instability, its eigenvalue tends
to zero as the resistivity $\eta \to 0$. Consequently,
the $\mathcal{O}(\sigma^2)$ terms on the left-hand side of
Eq.~(\ref{bz-relation}) can be neglected in the ideal outer region.

To eliminate $\delta B_y$ and $\delta\Delta p$, we use the ideal $y$-induction
equation together with Eq.~(\ref{eq:barbeta}). After substituting $\delta v_z$
and $\delta v_z'$, these two relations can be written as
\begin{align}
    \sigma \delta B_y &=
    ikB_0\delta v_y+\frac{i\sigma}{k}\frac{B_g'}{B_0}\delta B_z, \\
    \sigma \delta\Delta p &=
    -\frac{i}{k}\bar{\beta}B_0
    \left[
    \sigma\left(\delta B_z'-\frac{B_0'}{B_0}\delta B_z\right)
    +k^2B_g\delta v_y
    \right].
\end{align}
Solving the resulting ideal algebraic system for $\delta v_y$, $\delta B_y$,
and $\delta\Delta p$, and retaining the leading order in $\sigma^2/(k^2B_0^2)$,
yields
\begin{widetext}
\begin{align}
    \delta v_y &=
    -\frac{\sigma}{k^2B_0}
    \frac{\left(\bar{\beta}+\Delta\beta_0\right)B_g}{\mathcal{R}}
    \left(B_0\delta B_z'-B_0'\delta B_z\right), \\
    \delta B_y &=
    -\frac{i}{kB_0}
    \frac{
    \left(\bar{\beta}+\Delta\beta_0\right)B_0^2B_g\delta B_z'
    -\left(1-\frac{1}{2}\Delta\beta_0\right)B_g'\delta B_z
    }{\mathcal{R}}, \\
    \delta\Delta p &=
    -\frac{i}{k}\bar{\beta}
    \frac{1-\frac{1}{2}\Delta\beta_0}{\mathcal{R}}
    \left(B_0\delta B_z'-B_0'\delta B_z\right),
\end{align}
\end{widetext}
where
\begin{equation}
    \mathcal{R}(z)
    \equiv
    1-\frac{1}{2}\Delta\beta_0
    +\left(\bar{\beta}+\Delta\beta_0\right)B_g^2 .
\end{equation}

The final step is to substitute these expressions and their derivatives back
into the quasistatic form of Eq.~(\ref{bz-relation}). Using the equilibrium
identities listed in Appendix~\ref{app:equilibrium-identities}, the equation
reduces to
\begin{align}
  \label{eq:outer-explicit}
  \delta B_z''
  + \mathcal{P}(z)\delta B_z'
  + \mathcal{Q}(z)\delta B_z = 0,
\end{align}
with
\begin{align}
    \mathcal{P}(z) &=
    \frac{2\left(\bar{\beta}+\Delta\beta_0\right)B_0B_g^2}{\mathcal{R}}, \\
    \mathcal{Q}(z) &=
    \frac{2\left(1-\frac{1}{2}\Delta\beta_0\right)B_g^2}{\mathcal{R}}
    -k^2
    \frac{\mathcal{R}}{1+\bar{\beta}+\frac{1}{2}\Delta\beta_0}.
\end{align}
For $\Delta\beta_0=0$, $\bar{\beta}$ reduces to
\begin{equation}
    \tilde{\beta}
    \equiv
    \frac{1}{2}\left(\gamma_\parallel+\gamma_\perp-2\right)\beta_0,
\end{equation}
and Eq.~(\ref{eq:outer-explicit}) recovers the isotropic-equilibrium result of
\cite{2025ApJ...993...74F}. In the further limit $\bar{\beta}=0$ and
$\Delta\beta_0=0$, it reduces to the classical MHD outer-region equation.

More explicitly, in the asymptotic region $z \gg 1$, we have
$B_0 \to 1$, $B_g \to 0$, and all derivatives of the equilibrium field
vanish exponentially. It is useful to define
\begin{equation}
  \mathcal{A}\equiv 1-\frac{1}{2}\Delta\beta_0,
  \qquad
  \mathcal{R}_0\equiv
  1+\bar{\beta}+\frac{1}{2}\Delta\beta_0 .
  \label{eq:outer_A_R0}
\end{equation}
Using Eq.~(\ref{eq:barbeta}), the second quantity can be written as
\begin{equation}
  \mathcal{R}_0
  =
  1+\frac{1}{2}\left(\gamma_\parallel+\gamma_\perp-2\right)\beta_0
  +\frac{1}{2}\gamma_\parallel\Delta\beta_0 .
\end{equation}
Although the quasistatic outer equation is obtained by neglecting terms of
order $\sigma^2$, these terms should be retained in the far-field decay
condition when $\mathcal{A}$ becomes small, because then
$k^2\mathcal{A}$ can be comparable to $\sigma^2$. Keeping the finite-eigenvalue
terms in the large-$z$ limit gives
\begin{equation}
  \left(\sigma^2+k^2\mathcal{R}_0\right)\delta B_z''
  -k^2\left(\sigma^2+k^2\mathcal{A}\right)\delta B_z=0 .
  \label{eq:outer-asymptotic-full}
\end{equation}
Therefore,
\begin{equation}
  \delta B_z''-\lambda_\infty^2\delta B_z=0,
  \qquad
  \lambda_\infty^2
  =
  k^2
  \frac{\sigma^2+k^2\mathcal{A}}
  {\sigma^2+k^2\mathcal{R}_0}.
  \label{eq:lambda-full}
\end{equation}
The localized solution for $z\to+\infty$ is then
$\delta B_z\propto e^{-\lambda_\infty z}$, provided
$\lambda_\infty^2>0$. The same decay rate applies for $z\ll -1$, since
$B_g\to0$ and only the sign of $B_0$ changes.
The condition $\lambda_\infty^2>0$ is therefore an additional localization
requirement for the tearing eigenmode. If the numerator and denominator in
Eq.~(\ref{eq:lambda-full}) have opposite signs, then
$\lambda_\infty^2<0$ and the far-field solutions become oscillatory rather
than exponentially decaying. Such behavior is incompatible with a localized
tearing perturbation satisfying $\delta B_z\to0$ as $|z|\to\infty$. Within
the present boundary-value formulation, parameter regimes for which
$\lambda_\infty^2\leq0$ are therefore excluded from the unstable tearing
spectrum and are interpreted as completely stabilized against localized
tearing modes.
For the unstable, non-oscillatory tearing branch used in the asymptotic
estimates, $\sigma=\gamma$ and this condition can be written explicitly by
introducing
\begin{equation}
    \chi\equiv\frac{\gamma^2}{k^2}\geq0 .
\end{equation}
Since $k^2>0$, Eq.~(\ref{eq:lambda-full}) requires
\begin{equation}
    \left(\chi+\mathcal{A}\right)
    \left(\chi+\mathcal{R}_0\right)>0 .
\end{equation}
Substituting the definitions of $\mathcal{A}$ and $\mathcal{R}_0$, the
localized outer solution exists only when the general condition
\begin{equation}
\begin{aligned}
    &\left(1+\chi-\frac{\Delta\beta_0}{2}\right)
    \Biggl[
    1+\chi
    +\frac{1}{2}\left(\gamma_\parallel+\gamma_\perp-2\right)\beta_0
    \\
    &\qquad
    +\frac{1}{2}\gamma_\parallel\Delta\beta_0
    \Biggr]>0 .
\end{aligned}
    \label{eq:lambda-localization-condition}
\end{equation}
is satisfied, together with the pressure constraints $\beta_0>0$ and
$\Delta\beta_0>-\beta_0$. This form is valid for arbitrary
$\gamma_\parallel$ and $\gamma_\perp$; the specific closures used in the
numerical calculations are introduced only in
Section~\ref{sec:numerical-results}. In the quasistatic limit $\chi\to0$,
one loss-of-localization boundary is $\Delta\beta_0=2$, where the effective
anisotropic tension factor $\mathcal{A}$ vanishes.
This upper boundary is directly related to the parallel firehose threshold.
In the present normalization,
$\beta_{\perp,0}=\beta_0$ and
$\beta_{\parallel,0}=\beta_0+\Delta\beta_0$, so the firehose marginal
condition $\beta_{\parallel,0}-\beta_{\perp,0}=2$ is simply
$\Delta\beta_0=2$. Thus, in the quasistatic limit, the loss of far-field
localization occurs at the same point at which the equilibrium magnetic
tension changes sign. On the opposite side, a perpendicular-pressure-dominated
equilibrium is subject to the mirror threshold
$\beta_{\perp,0}(\beta_{\perp,0}/\beta_{\parallel,0}-1)=1$, which gives
\begin{equation}
    \Delta\beta_0=-\frac{\beta_0}{1+\beta_0}.
\end{equation}
More negative values of $\Delta\beta_0$ are mirror unstable, even if the
large-scale tearing boundary-value problem still admits an exponentially
localized outer solution. Therefore, when the equilibrium is required to be
stable to the usual anisotropy-driven microinstabilities, the physically
relevant interval is further restricted approximately to
$-\beta_0/(1+\beta_0)<\Delta\beta_0<2$, with the finite-eigenvalue correction
to the upper localization boundary given by
$\Delta\beta_0=2(1+\chi)$.

In the strict quasistatic ordering, where
$\sigma^2\ll k^2\mathcal{A}$ and $\sigma^2\ll k^2\mathcal{R}_0$,
Eq.~(\ref{eq:lambda-full}) reduces to
\begin{equation}
  \delta B_z'' - \lambda^2 \delta B_z = 0,
  \qquad
  \lambda^2 =
  k^2\frac{\mathcal{A}}{\mathcal{R}_0}
  =
  k^2\frac{1-\frac{1}{2}\Delta\beta_0}
  {1+\bar{\beta}+\frac{1}{2}\Delta\beta_0}.
  \label{eq:lambda-general}
\end{equation}
This is the decay rate associated with the reduced ideal outer equation,
Eq.~(\ref{eq:outer-explicit}), and it is the value used in the local
analytic construction below. Equation~(\ref{eq:lambda-full}) should instead
be used for the asymptotic boundary condition whenever the finite-eigenvalue
contribution is comparable to the anisotropic-tension term. Therefore, the
imposed equilibrium anisotropy modifies the outer decay length through both
the numerator, which contains the equilibrium anisotropic tension, and the
denominator, which contains the pressure-anisotropy response.

Because the equilibrium fields satisfy $B_0(z)$ antisymmetric and $B_g(z)$
symmetric, the tearing perturbation $\delta B_z$ inherits a symmetric
structure with respect to $z$. Consequently, $\delta B_z(z)$ is continuous
across the midplane, but its derivative $\delta B'_z(z)$ exhibits opposite
signs on either side, producing the well-known jump condition that defines
the tearing stability parameter $\Delta'$. For this reason, it is sufficient
to analyze the region $z>0$, with the solution for $z<0$ obtained by
reflection symmetry. To incorporate the quasistatic asymptotic decay of
the reduced outer equation directly into the solution, we introduce the ansatz
\begin{equation}
\label{eq:ansatz}
  \delta B_z(z) = e^{-\lambda z} f(z),
\end{equation}
where $f(z)$ approaches a constant as $z \to \infty$, and $\lambda$ is given
by Eq.~(\ref{eq:lambda-general}). If the finite-eigenvalue correction in
Eq.~(\ref{eq:lambda-full}) is retained in the far-field boundary condition,
the same construction applies with $\lambda$ replaced by $\lambda_\infty$.

By substituting the ansatz (Eq.~\ref{eq:ansatz}) together with its first- and
second-order derivatives into Eq.~(\ref{eq:outer-explicit}), the common
exponential factor $e^{-\lambda z}$ cancels out. The derivatives are
\begin{align}
    \delta B_z' &= e^{-\lambda z}\left(f'-\lambda f\right), \\
    \delta B_z'' &= e^{-\lambda z}\left(f''-2\lambda f'+\lambda^2 f\right).
\end{align}
Therefore, the differential equation for $f(z)$ can first be written in the compact form
\begin{equation}
    f''+\left[\mathcal{P}(z)-2\lambda\right]f'
    +\left[\lambda^2-\lambda\mathcal{P}(z)+\mathcal{Q}(z)\right]f=0.
\end{equation}
Using Eq.~(\ref{eq:lambda-general}) to eliminate the asymptotic contribution,
and substituting the expressions for $\mathcal{P}(z)$ and $\mathcal{Q}(z)$, we obtain
\begin{widetext}
\begin{align}
\label{eq:f_ode}
    f''(z)
    & +2\left[
    \frac{\left(\bar{\beta}+\Delta\beta_0\right)B_0B_g^2}{\mathcal{R}}
    -\lambda
    \right]f'(z) \\
    & +B_g^2\left[
    \frac{2\left(1-\frac{1}{2}\Delta\beta_0\right)}{\mathcal{R}}
    -\frac{k^2\left(\bar{\beta}+\Delta\beta_0\right)}
    {1+\bar{\beta}+\frac{1}{2}\Delta\beta_0}
    -\frac{2\lambda\left(\bar{\beta}+\Delta\beta_0\right)B_0}{\mathcal{R}}
    \right]f(z)=0. \nonumber
\end{align}
\end{widetext}

In general, this second-order differential equation has no closed-form analytic
solution. A local expansion near the sheet center does not independently
determine $\Delta'$, because the local coefficients remain constrained by the
global decaying outer solution. It does, however, express $\Delta'$ in terms of
the central curvature ratio $a_2/a_0$ and provide analytic constraints on that
ratio. We therefore develop the local series below, and subsequently determine
its required global input from exact limiting solutions and numerical
integration of the outer boundary-value problem.

\subsection{Derivation of the Tearing Stability Index \texorpdfstring{$\Delta'$}{Delta Prime}}
\label{ssec:jump}

To obtain the tearing stability parameter $\Delta'$ in the gyrotropic MHD framework,
we first introduce the dimensionless coordinate $\zeta=z/a$ and then transform
Eq.~(\ref{eq:f_ode}) into a more convenient local form by setting $u=\tanh\zeta$,
which implies $\sech^2\zeta = 1 - u^2$. In this local construction, $k$, $\lambda$,
and $\Delta'$ should be read as the dimensionless combinations $ka$, $a\lambda$, and
the sheet-thickness-normalized jump defined above, respectively. Using $\mathcal{A}$
from Eq.~(\ref{eq:outer_A_R0}), we define
\begin{align}
    \mu &\equiv \bar{\beta}+\Delta\beta_0, \\
    \mathcal{R}_u(u)&\equiv \mathcal{A}+\mu(1-u^2).
\end{align}
Then $\mathcal{R}_u(0)=\mathcal{R}_0$, with $\mathcal{R}_0$ as defined in
Eq.~(\ref{eq:outer_A_R0}). In this notation,
Eq.~(\ref{eq:lambda-general}) becomes
$(a\lambda)^2=(ka)^2\mathcal{A}/\mathcal{R}_0$, or
$\lambda^2=k^2\mathcal{A}/\mathcal{R}_0$ in the dimensionless variables used
in the series expansion.

The derivatives with respect to $\zeta$ transform as $f'(\zeta) = (1-u^2) f'(u)$
and $f''(\zeta) = (1-u^2)^2 f''(u) - 2u(1-u^2)f'(u)$. Substituting these into
Eq.~(\ref{eq:f_ode}) and dividing by the common factor $(1-u^2)$ gives
the equivalent differential equation for $f(u)$:
\begin{align}
\label{eq:f_ode_u}
& \mathcal{R}_u(1-u^2) f''(u)
 + \left[2\mu u(1-u^2)-2(\lambda+u)\mathcal{R}_u\right] f'(u) \\
& + \left[
2\mathcal{A}
-\frac{k^2\mu}{\mathcal{R}_0}\mathcal{R}_u
-2\lambda\mu u
\right] f(u)=0. \nonumber
\end{align}

Because the tearing solution $\delta B_z(\zeta)$ is symmetric, its derivative
with respect to $\zeta$ is antisymmetric. This property simplifies the jump
condition, $\Delta'$, to
\begin{equation}
  \Delta' =
  \frac{2}{\delta B_z(0)}
  \left.\frac{d\delta B_z}{d\zeta}\right|_{0^+}.
\end{equation}
Using the dimensionless form of the ansatz,
$\delta B_z(\zeta)=e^{-\lambda\zeta}f(\zeta)$, we evaluate this expression
at $\zeta=0$ and obtain
\begin{equation}
  \Delta' = 2 \left[ \frac{f'(0^+)}{f(0)} - \lambda \right].
\end{equation}
Thus, the tearing parameter is determined by the slope-to-value ratio of
$f$ at the sheet center, corrected by the decay rate $\lambda$.

To compute this ratio, we expand $f(u)$ in a power series around $u=0$:
\begin{equation}
    f(u) = \sum_{m=0}^{\infty} a_m u^m .
    \label{series}
\end{equation}
At the origin, $f(0)=a_0$ and $f'(0)=a_1$, so
\begin{equation}
    \Delta' = 2\left( \frac{a_1}{a_0} - \lambda \right).
    \label{eq:jump_approx}
\end{equation}
Here $f'(0)$ in the series expansion denotes $df/du$ at $u=0$; this is equal
to $df/d\zeta$ at the sheet center because $du/d\zeta=\sech^2\zeta=1$ at $\zeta=0$.

The coefficients of the series are constrained by Eq.~(\ref{eq:f_ode_u}).
Evaluating the equation at $u=0$ gives
\begin{equation}
    2\mathcal{R}_0a_{2}
    -2\lambda\mathcal{R}_0a_{1}
    +\left(2\mathcal{A}-k^2\mu\right)a_{0}=0.
\label{eq:coeffs_relation}
\end{equation}
From this relation,
\begin{equation}
    \frac{a_{1}}{a_{0}}=
    \left(\frac{\mathcal{A}}{\mathcal{R}_0}+\frac{a_{2}}{a_{0}}\right)\frac{1}{\lambda}
    -\frac{\mu}{2\mathcal{A}}\lambda .
\end{equation}
Substituting into Eq.~(\ref{eq:jump_approx}) yields the general form of
the tearing parameter,
\begin{equation}
    \Delta' =
    2\left( \frac{\mathcal{A}}{\mathcal{R}_0} + \frac{a_2}{a_0} \right)\frac{1}{\lambda}
    -\left(2+\frac{\mu}{\mathcal{A}}\right)\lambda ,
\label{eq:Delta_full}
\end{equation}
which shows that $\Delta'$ depends explicitly on the equilibrium pressure anisotropy
through $\mathcal{A}$, $\mu$, and $\mathcal{R}_0$, as well as on the ratio $a_2/a_0$
that encodes the curvature of the solution at the sheet center.

The series expansion in Eq.~(\ref{series}) generates a recursive relation for all
higher-order coefficients $a_m$. Collecting the coefficient of $u^n$ in
Eq.~(\ref{eq:f_ode_u}), and using $\lambda^2=k^2\mathcal{A}/\mathcal{R}_0$, gives
\begin{widetext}
\begin{align}
&\mathcal{R}_0(n+2)(n+1)a_{n+2}
-2\lambda\mathcal{R}_0(n+1)a_{n+1} \nonumber\\
&+\left[
2\mathcal{A}(1-n)-(\mathcal{R}_0+\mu)n(n-1)
-\frac{\mathcal{R}_0\mu}{\mathcal{A}}\lambda^2
\right]a_n \nonumber\\
&+2\lambda\mu(n-2)a_{n-1}
+\left[
\mu(n-2)(n-3)+\frac{\mu^2}{\mathcal{A}}\lambda^2
\right]a_{n-2}=0 ,
\label{eq:am_recursion}
\end{align}
where $n\geq0$ and $a_{-1}=a_{-2}=0$. Equivalently, Eq.~(\ref{eq:am_recursion})
can be solved for $a_{n+2}$:
\begin{align}
a_{n+2} &=
\frac{2\lambda}{n+2}a_{n+1}
-\frac{1}{\mathcal{R}_0(n+2)(n+1)}
\Biggl\{
\left[
2\mathcal{A}(1-n)-(\mathcal{R}_0+\mu)n(n-1)
-\frac{\mathcal{R}_0\mu}{\mathcal{A}}\lambda^2
\right]a_n \nonumber\\
&\hspace{1.3in}
+2\lambda\mu(n-2)a_{n-1}
+\left[
\mu(n-2)(n-3)+\frac{\mu^2}{\mathcal{A}}\lambda^2
\right]a_{n-2}
\Biggr\}.
\end{align}
\end{widetext}

The exact marginal condition can be obtained from the global outer equation.
Restoring the dimensionless wavenumber $\alpha\equiv ka$, the
marginal point is
\begin{equation}
    \alpha_c^2=\frac{\mathcal{A}}{\mathcal{R}_0}.
    \label{eq:alpha_marginal}
\end{equation}
At this value the outer equation admits the exact even solution
\begin{equation}
    b_m(\zeta)=\sech^{p}\zeta,
    \qquad
    p\equiv\frac{\mathcal{A}}{\mathcal{R}_0},
    \label{eq:exact_marginal_outer}
\end{equation}
where $b\equiv\delta B_z$. Since $b_m'(0)=0$, this solution has
$\Delta'=0$. Its far-field decay is also consistent with the asymptotic
condition because $\lambda a=\alpha_c\sqrt{\mathcal{A}/\mathcal{R}_0}=p$.
Thus, for the localized branch considered here, the global marginal condition
is $\alpha=\alpha_c$, and tearing-unstable modes satisfy
\begin{equation}
    ka < \sqrt{\frac{\mathcal{A}}{\mathcal{R}_0}} .
    \label{eq:global_tearing_condition}
\end{equation}
In the classical MHD limit $\beta_0\to0$ and $\Delta\beta_0\to0$, where
$\mathcal{A}=\mathcal{R}_0=1$, this reduces to the standard condition
$ka<1$.
A direct numerical integration of the outer-region equation, described in
Appendix~\ref{app:outer-stability-boundary}, confirms that the zero of
$\Delta'$ occurs at Eq.~(\ref{eq:alpha_marginal}) for the three closures
used below.

This exact marginal solution also provides a useful global constraint on the
local series coefficient. For the transformed function
$b=e^{-\lambda a\zeta}f$, Eq.~(\ref{eq:exact_marginal_outer}) gives, at
marginality,
\begin{equation}
    f_m=e^{p\zeta}\sech^p\zeta=(1+u)^p .
\end{equation}
Therefore
\begin{equation}
    \left.\frac{a_2}{a_0}\right|_{\alpha=\alpha_c}
    =
    \frac{p(p-1)}{2}
    =
    -\frac{\mathcal{A}\mu}{2\mathcal{R}_0^2}.
    \label{eq:a2a0_marginal_exact}
\end{equation}
Together with the long-wavelength requirement $a_2/a_0\to0$ as
$\alpha\to0$, this suggests the globally constrained approximation
\begin{equation}
    \frac{a_2}{a_0}
    \approx
    -\frac{\mu}{2\mathcal{R}_0}\alpha^2
    =
    -\frac{\mu}{2\mathcal{A}}(\lambda a)^2 .
    \label{eq:a2a0_global_closure}
\end{equation}
Substitution into Eq.~(\ref{eq:Delta_full}) gives
\begin{equation}
    \Delta'
    \approx
    2\left[
    \frac{1}{\alpha}\sqrt{\frac{\mathcal{A}}{\mathcal{R}_0}}
    -
    \alpha\sqrt{\frac{\mathcal{R}_0}{\mathcal{A}}}
    \right],
    \label{eq:Delta_global_closure}
\end{equation}
which preserves both the exact long-wavelength coefficient and the exact
marginal point in Eq.~(\ref{eq:alpha_marginal}). This closure for
$a_2/a_0$ is nevertheless a limited interpolation between two global
constraints, not a full solution of the outer boundary-value problem.
Consequently, Eq.~(\ref{eq:Delta_global_closure}) should be interpreted as a
compact approximation to the true $\Delta'$, useful mainly because it enforces
the correct leading long-wavelength behavior and the correct zero at
marginality. In Appendix~\ref{app:outer-stability-boundary}, we construct a
closer empirical expression for $a_2/a_0$ that still preserves the exact
marginal condition and gives a more accurate tearing index,
$\Delta'_{\rm fit}$, constructed from the fitted curvature ratio for
comparison with the numerical dispersion relations.

\subsection{Inner Layer Solution}
\label{ssec:inner}

We now consider the resistive inner layer around the resonant surface. Let $\delta$
denote the characteristic layer width and introduce the stretched coordinate
$\xi=z/\delta$, with $\delta\ll a$. Near $z=0$, the equilibrium field can be expanded as
\begin{align}
    B_0(z)&=B_0'(0)z+\mathcal{O}(z^3),
    &
    B_g(z)&=1+\mathcal{O}(z^2), \\
    B_g'(z)&=\mathcal{O}(z),
    &
    B_0''(z)&=\mathcal{O}(z),
\end{align}
where $B_0'(0)=1/a$ for $B_0=\tanh(z/a)$.

Substitution of these expansions into the $z$-momentum and $z$-induction equations gives,
to the order needed for the inner-layer balance,
\begin{widetext}
\begin{align}
    \sigma\left(\delta v_z''-k^2\delta v_z\right)
    &=
    i k \mathcal{A} B_0'(0) z
    \left(\delta B_z''-k^2\delta B_z\right)
    -2k^2 B_0'^2(0) z\,\delta\Delta p \nonumber\\
    &\quad
    +\nu\left(\delta v_z''''-2k^2\delta v_z''+k^4\delta v_z\right)
    +\mathcal{O}(z^2),
    \label{eq:inner_vz_full}\\
    \sigma\delta B_z
    &=
    i k B_0'(0)z\,\delta v_z
    +\eta\left(\delta B_z''-k^2\delta B_z\right).
    \label{eq:inner_bz_full}
\end{align}
\end{widetext}
Here $\mathcal{A}$ is the effective tension factor introduced in
Eq.~(\ref{eq:outer_A_R0}). The term proportional to $\mathcal{A}$ represents
the modification of magnetic tension by the imposed equilibrium pressure
anisotropy.

For the standard tearing ordering, $k\delta\ll1$, derivatives across the inner layer
dominate over variations along the current sheet:
\begin{equation}
    \partial_z^2 \sim \delta^{-2},
    \qquad
    k^2 \ll \delta^{-2}.
\end{equation}
Under this ordering, and in the inviscid limit $\nu=0$, the pressure term in
Eq.~(\ref{eq:inner_vz_full}) is subdominant in the reconnecting-field balance.

This can be seen directly from the pressure-difference equation,
Eq.~(\ref{dp_lin}). In the inner layer, $B_0\sim z/a\sim\hat{\delta}$,
$B_g\sim1$, and the leading induction balance gives
$\gamma\tau_A\,\delta B_z\sim\alpha\hat{\delta}\,\delta v_z
\sim S^{-1}\delta B_z''$, where $\hat{\delta}\equiv\delta/a$ and, in this
estimate, primes on $\delta B_z$ denote derivatives with respect to $z/a$.
Both the field-line-straining source and the Ohmic source in
Eq.~(\ref{dp_lin}) then give the conservative estimate
\begin{equation}
    \delta\Delta p
    \sim
    \frac{\mathcal{C}_p}{\alpha}\,\delta B_z ,
    \qquad
    \mathcal{C}_p=O\!\left(|\bar{\beta}|+|\gamma_\parallel-1|\right),
    \label{eq:inner_deltap_estimate}
\end{equation}
where $\mathcal{C}_p$ is independent of $S$ and vanishes for the
double-isothermal pressure-difference response. The ratio of the pressure
force retained in Eq.~(\ref{eq:inner_vz_full}) to the magnetic-tension term is
therefore
\begin{equation}
    \epsilon_p
    \equiv
    \frac{\left|k^2B_0'^2(0)z\,\delta\Delta p\right|}
    {\left|k\mathcal{A}B_0'(0)z\,\delta B_z''\right|}
    \sim
    \frac{\mathcal{C}_p}{\mathcal{A}}
    \frac{\delta B_z}{\delta B_z''}.
    \label{eq:inner_pressure_ratio}
\end{equation}
For the constant-$\psi$ FKR ordering,
$\delta B_z''\sim(\Delta'/\hat{\delta})\delta B_z$, so, using the FKR layer
width obtained below in Eq.~(\ref{eq:fkr_delta}),
\begin{equation}
    \epsilon_p^{\rm FKR}
    \sim
    \frac{\mathcal{C}_p}{\mathcal{A}}\frac{\hat{\delta}}{\Delta'}
    \sim
    \frac{\mathcal{C}_p}{\mathcal{A}^{6/5}}
    \alpha^{-2/5}\left(\Delta'\right)^{-4/5}S^{-2/5}.
    \label{eq:inner_pressure_ratio_fkr}
\end{equation}
Thus the pressure-fluctuation force is subdominant throughout the asymptotic
FKR interval for fixed $\alpha$ and finite $\Delta'$, apart from the immediate
marginal neighborhood where $\Delta'\to0$ and the FKR expansion is not uniform.
For the nonconstant-$\psi$ Coppi ordering,
$\delta B_z''\sim\delta B_z/\hat{\delta}^2$, giving, with
Eq.~(\ref{eq:coppi_delta}),
\begin{equation}
    \epsilon_p^{\rm Coppi}
    \sim
    \frac{\mathcal{C}_p}{\mathcal{A}}\hat{\delta}^2
    \sim
    \frac{\mathcal{C}_p}{\mathcal{A}^{4/3}}
    \alpha^{-2/3}S^{-2/3}\ll1 .
    \label{eq:inner_pressure_ratio_coppi}
\end{equation}
The $\delta\Delta p$ term therefore supplies only a higher-order correction to
the reconnecting-field balance used below, while the imposed equilibrium
anisotropy enters at leading order through the effective tension
$\mathcal{A}$.

Because the estimates contain negative powers of
$\mathcal{A}$ and, in the FKR ordering, $\Delta'$, they should not be
extrapolated arbitrarily close to the firehose/localization boundary
$\mathcal{A}\to0$ or to the marginal limit $\Delta'\to0$.
The leading inner equations reduce to
\begin{align}
    \sigma\delta v_z'' &=
    i k \mathcal{A} B_0'(0)z\,\delta B_z'',
    \label{eq:inner_vz}\\
    \sigma\delta B_z &=
    i k B_0'(0)z\,\delta v_z
    +\eta\delta B_z''.
    \label{eq:inner_bz}
\end{align}
In terms of the stretched coordinate $\xi$, these become
\begin{align}
    \sigma\frac{\partial^2\delta v_z}{\partial \xi^2}
    &=
    i k \mathcal{A} B_0'(0)\delta\,\xi
    \frac{\partial^2\delta B_z}{\partial \xi^2},
    \label{eq:inner_vz_xi}\\
    \sigma\delta B_z
    &=
    i k B_0'(0)\delta\,\xi\,\delta v_z
    +\frac{\eta}{\delta^2}
    \frac{\partial^2\delta B_z}{\partial \xi^2}.
    \label{eq:inner_bz_xi}
\end{align}

Equations~(\ref{eq:inner_vz})--(\ref{eq:inner_bz}), or equivalently
Eqs.~(\ref{eq:inner_vz_xi})--(\ref{eq:inner_bz_xi}), are the starting point
for the inner-layer term balances. Under the same ordering, the remaining
equations for $\delta v_y$, $\delta B_y$, and $\delta\Delta p$ are determined
after the leading reconnecting subsystem is solved; they describe the local
pressure-anisotropy response but do not modify the dominant inner-layer
dynamics at this order.

This represents an important distinction from the isotropic-equilibrium case
studied by \citet{2025ApJ...993...74F}. There, the gyrotropic corrections
entered the growth-rate scalings through the outer-region matching parameter
$\Delta'$, and therefore affected primarily the constant-$\psi$ branch. Here,
the prescribed equilibrium pressure anisotropy appears explicitly in the leading
inner-layer momentum balance through the factor $\mathcal{A}$. Consequently,
both the constant-$\psi$ and nonconstant-$\psi$ regimes are expected to inherit
a direct dependence on the equilibrium anisotropy.

\subsection{Inner-Layer Term Balances and Growth-Rate Scalings}
\label{ssec:inner-scalings}

We now use Eqs.~(\ref{eq:inner_vz})--(\ref{eq:inner_bz}) to estimate the tearing
growth rate and inner-layer thickness.
We introduce the dimensionless quantities
\begin{equation}
    \alpha\equiv k a,
    \qquad
    \hat{\delta}\equiv\frac{\delta}{a}.
\end{equation}
The Lundquist number $S$ and Alfv\'en crossing time $\tau_A$ are defined in
Section~\ref{sec:equations}.
Here $\Delta'$ is the same dimensionless outer tearing index defined in
Eq.~(\ref{eq:jump}). Because we retain only the asymptotic
parameter dependences, dimensionless numerical coefficients of order unity are
omitted from the scaling relations below. Expressing the unstable-mode growth
rate as $\gamma={\rm Re}(\sigma)$, the leading inner equations imply the balances
used below for the unstable nonoscillatory tearing branch, for which
$\mathcal{A}>0$ and $\mathcal{R}_0>0$. The weaker condition
$\mathcal{A}/\mathcal{R}_0>0$ ensures a real quasistatic outer decay rate but
also permits both coefficients to be negative; that regime is outside the
inner-layer growth-rate scalings derived here.
\begin{align}
    \gamma\tau_A\,\frac{\delta v_z}{\hat{\delta}^2}
    &\sim
    \mathcal{A}\alpha\hat{\delta}\,\delta B_z'', \\
    \gamma\tau_A\,\delta B_z
    &\sim
    \alpha\hat{\delta}\,\delta v_z
    \sim
    S^{-1}\delta B_z'' ,
\end{align}
where the double prime now denotes differentiation with respect to $z/a$.

In the constant-$\psi$ or FKR regime \citep{1963PhFl....6..459F}, the
reconnecting magnetic perturbation varies weakly across the layer, but its
derivative must match the outer solution. Therefore
\begin{equation}
    \delta B_z''\sim\frac{\Delta'}{\hat{\delta}}\delta B_z,
    \qquad
    \delta v_z''\sim\frac{\delta v_z}{\hat{\delta}^2}.
\end{equation}
The induction equation gives
\begin{equation}
    \gamma\tau_A\sim S^{-1}\frac{\Delta'}{\hat{\delta}},
\end{equation}
while combining the momentum equation with the convective part of induction gives
\begin{equation}
    \left(\gamma\tau_A\right)^2
    \sim
    \mathcal{A}\alpha^2\Delta'\hat{\delta}^3 .
\end{equation}
Eliminating $\hat{\delta}$ yields
\begin{align}
    \gamma\tau_A
    &\sim
    \mathcal{A}^{1/5}\alpha^{2/5}
    \left(\Delta'\right)^{4/5}S^{-3/5},
    \label{eq:fkr_growth}\\
    \frac{\delta}{a}
    &\sim
    \mathcal{A}^{-1/5}\alpha^{-2/5}
    \left(\Delta'\right)^{1/5}S^{-2/5}.
    \label{eq:fkr_delta}
\end{align}
Thus, in the FKR branch the equilibrium anisotropy affects the growth rate
both through the inner-layer coefficient $\mathcal{A}$ and through the outer
matching parameter $\Delta'$.

In the nonconstant-$\psi$ or Coppi regime \citep{1976SvJPP...2..533C}, the
reconnecting perturbation changes substantially within the inner layer, so the
appropriate estimate is
\begin{equation}
    \delta B_z''\sim\frac{\delta B_z}{\hat{\delta}^2}.
\end{equation}
The resistive part of the induction equation then gives
\begin{equation}
    \gamma\tau_A\sim S^{-1}\hat{\delta}^{-2},
\end{equation}
whereas the momentum-convection balance gives
\begin{equation}
    \left(\gamma\tau_A\right)^2
    \sim
    \mathcal{A}\alpha^2\hat{\delta}^2.
\end{equation}
Solving these two relations gives
\begin{align}
    \gamma\tau_A
    &\sim
    \mathcal{A}^{1/3}\alpha^{2/3}S^{-1/3},
    \label{eq:coppi_growth}\\
    \frac{\delta}{a}
    &\sim
    \mathcal{A}^{-1/6}\alpha^{-1/3}S^{-1/3}.
    \label{eq:coppi_delta}
\end{align}
Unlike the isotropic-equilibrium case, the Coppi branch now depends explicitly
on the prescribed equilibrium anisotropy through $\mathcal{A}$.

\subsection{Scaling of the Fastest-Growing Mode}
\label{ssec:max-growth-scalings}

The fastest-growing mode is obtained by matching the FKR and Coppi branches.
For this purpose, the relevant outer-region input is the global
long-wavelength limit of $\Delta'$. Setting $\alpha=0$ in the outer equation
gives the two independent zero-wavenumber solutions, with
$b\equiv\delta B_z$,
\begin{equation}
    b_1=\tanh\zeta,
    \qquad
    b_2=\mathcal{A}\zeta\tanh\zeta-\mathcal{R}_0 .
\end{equation}
In the overlap region $1\ll\zeta\ll(\lambda a)^{-1}$, the decaying solution
behaves as $e^{-\lambda a\zeta}\simeq1-\lambda a\zeta$. Matching this behavior
to the zero-wavenumber outer solution gives $b(0)\sim(\lambda a)\mathcal{R}_0/\mathcal{A}$
and $db/d\zeta|_0\sim1$. Therefore,
\begin{equation}
    \Delta'
    \sim
    \frac{2\mathcal{A}}{(\lambda a)\mathcal{R}_0}
    =
    \frac{2}{\alpha}\sqrt{\frac{\mathcal{A}}{\mathcal{R}_0}}
    \equiv
    \frac{\mathcal{C}_\Delta}{\alpha},
\end{equation}
with
\begin{equation}
    \mathcal{C}_\Delta
    \equiv
    2\sqrt{\frac{\mathcal{A}}{\mathcal{R}_0}},
    \label{eq:CDelta}
\end{equation}
for the unstable nonoscillatory localized branch with
$\mathcal{A}>0$ and $\mathcal{R}_0>0$.
Substituting this form into Eq.~(\ref{eq:fkr_growth}) and equating it to
Eq.~(\ref{eq:coppi_growth}) gives
\begin{equation}
    \alpha_{\max}
    \sim
    \mathcal{C}_\Delta^{3/4}
    \mathcal{A}^{-1/8}
    S^{-1/4}
    \sim
    \mathcal{A}^{1/4}\mathcal{R}_0^{-3/8}S^{-1/4}.
\end{equation}
The corresponding maximum growth rate and inner-layer thickness are
\begin{align}
    \gamma_{\max}\tau_A
    &\sim
    \mathcal{C}_\Delta^{1/2}
    \mathcal{A}^{1/4}
    S^{-1/2}
    \sim
    \mathcal{A}^{1/2}\mathcal{R}_0^{-1/4}S^{-1/2},
    \label{eq:gamma_max_scaling}\\
    \frac{\delta_{\max}}{a}
    &\sim
    \mathcal{C}_\Delta^{-1/4}
    \mathcal{A}^{-1/8}
    S^{-1/4}
    \sim
    \mathcal{A}^{-1/4}\mathcal{R}_0^{1/8}S^{-1/4}.
    \label{eq:delta_max_scaling}
\end{align}
When $\Delta\beta_0\to0$, $\mathcal{A}\to1$ and the direct
anisotropic-tension correction disappears from the inner-layer balances, while
the remaining $\mathcal{R}_0$ dependence is inherited from the outer solution.
If the pressure-response correction is also absent, $\mathcal{R}_0\to1$, the
classical maximum-mode estimates are recovered, with
$\gamma_{\max}\tau_A\sim S^{-1/2}$ and $\alpha_{\max}\sim S^{-1/4}$.

\subsection{Secondary Pressure-Anisotropy Scales}
\label{ssec:secondary-pressure-scales}

Applying the near-resonant-surface equilibrium-field expansion introduced in
Section~\ref{ssec:inner}, $B_0\simeq z/a$ and $B_g\simeq1$, to the coupled
$\delta v_y$ and $\delta\Delta p$ equations,
Eqs.~(\ref{dvy}) and (\ref{dp_lin}), exposes a secondary pressure-anisotropy
scale. At leading order, reducing this local two-field subsystem introduces
the stiffness denominator
$\sigma^2+k^2\bar{\beta}B_0^2B_g^2$, with $\bar{\beta}$ defined in
Eq.~(\ref{eq:barbeta}). For $\bar{\beta}>0$ this denominator is not singular,
but it varies rapidly across the central scale
\begin{equation}
    \frac{\delta_q}{a}
    \sim
    \frac{\gamma\tau_A}{(ka)\sqrt{\bar{\beta}}}.
    \label{eq:pressure_anisotropy_resolution_scale}
\end{equation}
Thus large-$\beta_0$ cases with a strong pressure response can develop a
narrow peak in $\delta\Delta p$ near $z=0$ even when the ordinary resistive
tearing layer is resolved. The derivative of this peak is also dynamically
relevant because the $z$-momentum equation contains the pressure force
$B_0(2B_0'\delta\Delta p+B_0\delta\Delta p')$. This scale does not change the
leading FKR and Coppi balances when the pressure force remains asymptotically
subdominant, but it is a distinct scale that must be resolved in the
eigenfunctions.

A complementary situation occurs whenever the pressure-response coefficient
is negative, but the parameter-space route into this regime is not unique.
For $\gamma_\parallel<1$,
\begin{equation}
    \bar{\beta}
    =
    \frac{1}{2}\left[
    \left(\gamma_\parallel+\gamma_\perp-2\right)\beta_0
    -\left(1-\gamma_\parallel\right)\Delta\beta_0
    \right],
\end{equation}
and therefore
\begin{equation}
    \bar{\beta}<0
    \quad\Longleftrightarrow\quad
    \left(\gamma_\parallel+\gamma_\perp-2\right)\beta_0
    <
    \left(1-\gamma_\parallel\right)\Delta\beta_0 .
\end{equation}
Equivalently,
\begin{equation}
    \Delta\beta_0
    >
    \frac{\gamma_\parallel+\gamma_\perp-2}{1-\gamma_\parallel}\,\beta_0 .
    \label{eq:negative_barbeta_condition}
\end{equation}
If $\gamma_\parallel+\gamma_\perp-2>0$, the threshold in
Eq.~(\ref{eq:negative_barbeta_condition}) is positive, so negative
$\bar{\beta}$ requires sufficiently large positive $\Delta\beta_0$ relative to
$\beta_0$. The double-polytropic closure is a simple example of this branch:
for
$(\gamma_\parallel,\gamma_\perp)=(0.5,2)$,
$\bar{\beta}=(\beta_0-\Delta\beta_0)/4$, and
Eq.~(\ref{eq:negative_barbeta_condition}) reduces to
$\Delta\beta_0>\beta_0$. If
$\gamma_\parallel+\gamma_\perp-2<0$, however, the threshold is negative. Then
negative $\bar{\beta}$ can also occur on the perpendicular-pressure-dominated
side, $\Delta\beta_0<0$, provided
\begin{equation}
    -\frac{\left|\gamma_\parallel+\gamma_\perp-2\right|}
    {1-\gamma_\parallel}\,\beta_0
    <
    \Delta\beta_0
    <
    0 .
\end{equation}
This negative-$\Delta\beta_0$ branch is distinct from the large-positive
anisotropy branch because the equilibrium anisotropy, pressure-positivity
constraint, and mirror-stability constraint are different, even though the
local $\delta v_y$--$\delta\Delta p$ block has the same sign change in
$\bar{\beta}$. For any negative-$\bar{\beta}$ case, the local denominator has
the form
\begin{equation}
    \mathcal{D}(z)
    \sim
    (\gamma\tau_A)^2
    -(ka)^2|\bar{\beta}|B_0^2B_g^2 .
\end{equation}
Since $B_0B_g$ vanishes at $z=0$ and reaches
$\max |B_0B_g|=1/2$ away from the sheet center, this denominator can become
small at off-center locations satisfying
\begin{equation}
    |B_0(z_j)B_g(z_j)|
    \sim
    \frac{\gamma\tau_A}{(ka)\sqrt{|\bar{\beta}|}} .
\end{equation}
Here $z_j$ labels these off-center locations, with the index $j$
distinguishing the distinct roots on either side of the current sheet.
Thus negative-$\bar{\beta}$ modes can develop pressure-anisotropy structure
away from the central tearing layer, and resolving only $z=0$ is then
insufficient for a converged eigenfunction.

\section{Numerical Results}
\label{sec:numerical-results}

\subsection{Numerical Method and Branch-Scaling Comparison}
\label{ssec:numerical-method}
\label{ssec:dispersion-branch-verification}

We solve the full linear eigenvalue problem with the Pseudo-Spectral
Eigenvalue Calculator with an Automated Solver (PSECAS) framework
\citep{2019MNRAS.485..908B}, following the same numerical strategy as in
\citet{2025ApJ...993...74F}. The perturbations are written with the normal-mode
dependence used in the analytical derivation, so the linearized equations
reduce to a coupled set of ordinary differential equations in $z$. PSECAS
discretizes these equations pseudospectrally and assembles the resulting
matrix problem in the form
\begin{equation}
    \mathbf{M}_1\mathbf{q}
    =
    \sigma\,\mathbf{M}_2\mathbf{q},
\end{equation}
where $\mathbf{q}$ is the vector of collocation values of the eigenfunctions
and $\sigma$ is the complex eigenvalue.

\begin{figure*}[!ht]
\centering
\includegraphics[width=0.47\textwidth]{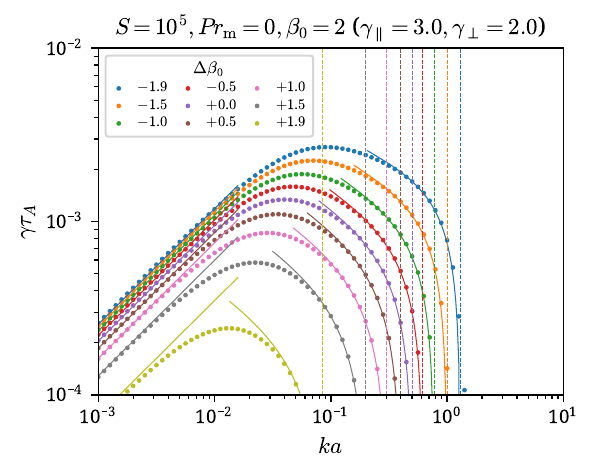}
\hfill
\includegraphics[width=0.47\textwidth]{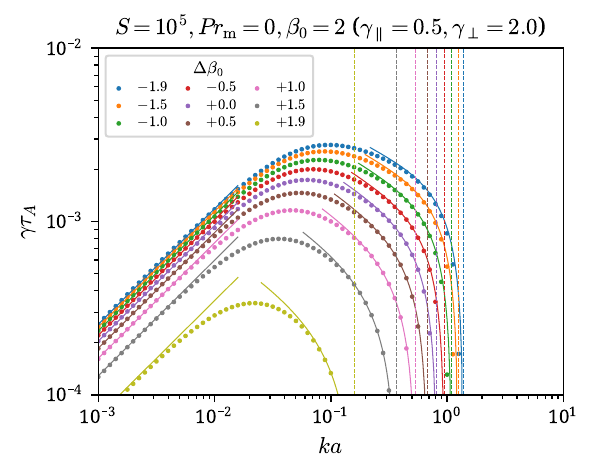}

\includegraphics[width=0.47\textwidth]{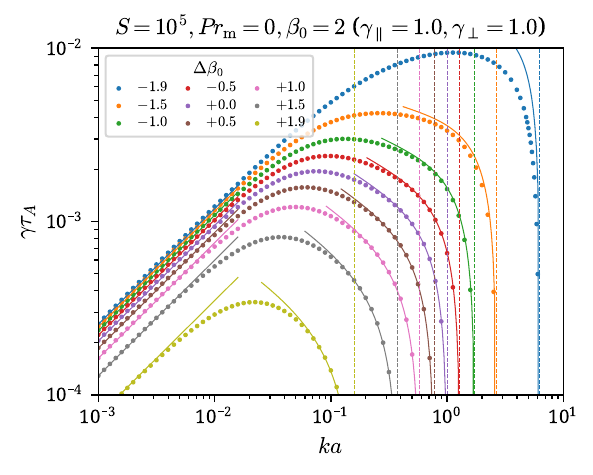}
\caption{Numerical dispersion relations and asymptotic branch scalings for
  $S=10^5$, $Pr_m=0$, and $\beta_0=2$. The upper-left, upper-right, and lower
  panels correspond to the double-adiabatic closure
  $(\gamma_\parallel,\gamma_\perp)=(3,2)$, the double-polytropic closure
  $(0.5,2)$, and the double-isothermal closure $(1,1)$, respectively. Symbols
  show growth rates calculated with PSECAS for different imposed anisotropies
  $\Delta\beta_0$, while solid curves show the theoretical Coppi and FKR branch
  scalings. The FKR curves use the fitted outer matching
  parameter $\Delta'_{\rm fit}$ constructed from the fitted curvature ratio
  $a_2/a_0$. The Coppi curves are independent of
  $\Delta'$ at leading order. The vertical dashed lines mark, for each
  anisotropy curve, the analytic marginal cutoff
$\alpha_c=\sqrt{\mathcal{A}/\mathcal{R}_0}$ from
Eq.~(\ref{eq:alpha_marginal}); values with $\alpha\geq\alpha_c$ are outside
the localized tearing-unstable interval.}
\label{fig:dispersion-branches}
\end{figure*}

The gyrotropic tearing equations were implemented as a dedicated PSECAS
system for the force-free equilibrium $B_x=\tanh(z/a)$ and
$B_y=\sech(z/a)$. The calculations reported here use two-dimensional tearing
modes with $k_x=k$ and $k_y=0$. We analyze three closures: the
double-adiabatic closure $(\gamma_\parallel,\gamma_\perp)=(3,2)$, the
double-polytropic closure $(\gamma_\parallel,\gamma_\perp)=(0.5,2)$, and the
double-isothermal closure $(\gamma_\parallel,\gamma_\perp)=(1,1)$. For the
non-isothermal closures, the solved
variables are
$(\delta v_z,\delta B_z,\delta v_y,\delta B_y,\delta\Delta p)$. In the
double-isothermal case, the pressure-anisotropy perturbation decouples and
the system reduces to
$(\delta v_y,\delta v_z,\delta B_y,\delta B_z)$. The coefficients in the
matrix problem include the resistivity $\eta=S^{-1}$, the viscosity
$\nu=Pr_m S^{-1}$, the prescribed equilibrium values $\beta_0$ and
$\Delta\beta_0$, and the closure indices
$(\gamma_\parallel,\gamma_\perp)$. At the two outer boundaries we impose the
asymptotic decaying form of the localized tearing eigenfunction,
$d_z f-\lambda f=0$ on the left boundary and
$d_z f+\lambda f=0$ on the right boundary, for each evolved variable $f$.
Here $\lambda$ is the quasistatic decay rate from
Eq.~(\ref{eq:lambda-general}). The exact far-field rate is
$\lambda_\infty$ from Eq.~(\ref{eq:lambda-full}), and the difference can be
largest in the small-wavenumber Coppi part of the dispersion relation. We
therefore repeated the most demanding case in the dispersion scan,
$(\gamma_\parallel,\gamma_\perp)=(1,1)$, $\beta_0=2$,
$\Delta\beta_0=-1.9$, $S=10^5$, and $Pr_m=0$, using
$\lambda_\infty$ in the outer boundary condition. This case has
$|\lambda_\infty/\lambda-1|$ of tens of percent at the smallest plotted
wavenumbers, but the growth rates were unchanged at the level of the
dispersion data: for $ka\leq0.1$, the maximum relative change in
${\rm Re}(\sigma)$ was below $7\times10^{-6}$, and over the common tabulated
wavenumbers it remained below $1.1\times10^{-3}$. This insensitivity is
expected because the rational Chebyshev grid places the effective outer
matching region deep in the exponential tail; for the small-$ka$ Coppi modes
in this test, the corresponding outer distance is hundreds of current-sheet
thicknesses. The quasistatic boundary condition therefore does not affect the
reported tearing growth rates within the numerical accuracy of the present
scan.

The spatial discretization uses the rational Chebyshev grid of PSECAS, which
maps the infinite domain to Chebyshev--Gauss collocation points through
$z=Cx/\sqrt{1-x^2}$. The scaling factor $C$ is chosen separately for each
wavenumber. It is constrained so that the current-sheet and resistive-layer
regions are resolved by a prescribed minimum number of collocation points,
while the outermost collocation points lie far enough into the exponential
tail that the amplitude is reduced to the specified fraction of its central
value. For the runs shown here, the grid selection requested 25 collocation
points within the current-sheet scale and, when an estimate of
$\delta_{\rm in}$ was available, at least five points within the resistive
layer. The outer amplitude fraction was typically set to 0.01 in the
dispersion scans, and the resolution was increased from $N=128$ in increments
of 32 up to the maximum needed for convergence.

For cases with $\bar{\beta}>0$, the grid selection must also resolve the
pressure-anisotropy scale $\delta_q$ from
Eq.~(\ref{eq:pressure_anisotropy_resolution_scale}) when it is smaller than,
or comparable to, the resistive layer. For $\bar{\beta}<0$, the analogous
requirement is to sample the off-center locations where
$(\gamma\tau_A)^2-(ka)^2|\bar{\beta}|B_0^2B_g^2$ becomes small. If either
scale is under-resolved, the $\delta v_y$ and $\delta\Delta p$ eigenfunctions
can develop grid-scale ringing, and the solver can reach the eigenvalue
tolerance on a nearby spurious root before the physical tearing eigenmode is
resolved.

\begin{figure*}[!ht]
\centering
\includegraphics[width=0.95\textwidth]{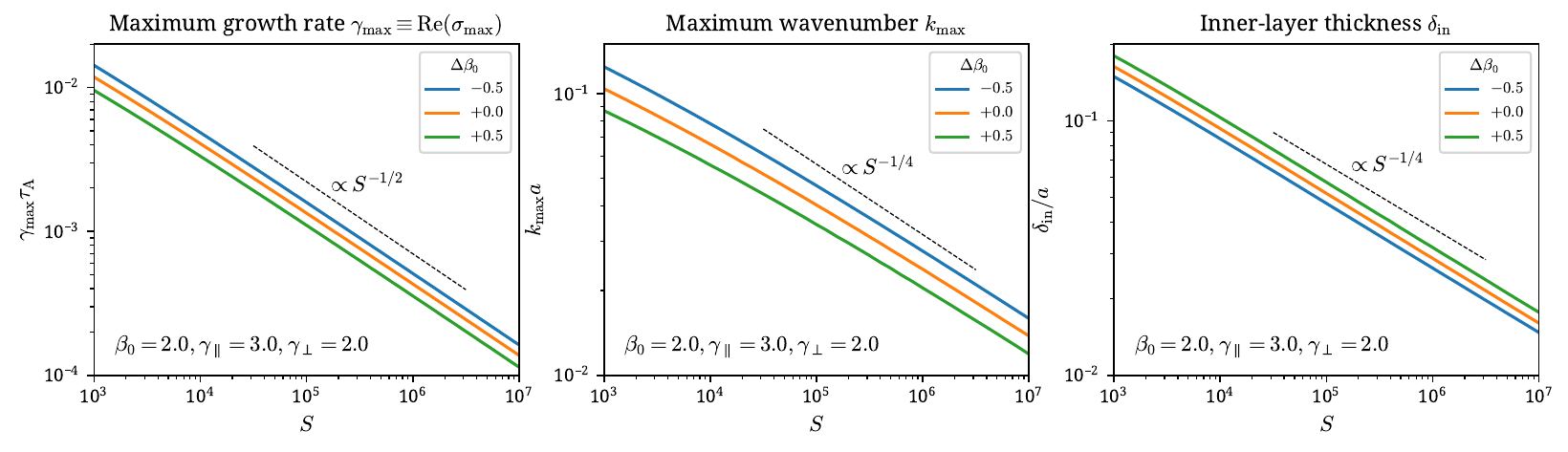}
\caption{Lundquist-number dependence of the fastest-growing tearing
mode for the double-adiabatic closure
$(\gamma_\parallel,\gamma_\perp)=(3,2)$, with $\beta_0=2$ and $Pr_m=0$. The
three curves correspond to $\Delta\beta_0=-0.5$, $0$, and $+0.5$. The panels
show, from left to right, the maximum growth rate, the corresponding
wavenumber, and the inner-layer thickness. Dashed lines indicate the
theoretical slopes $S^{-1/2}$, $S^{-1/4}$, and $S^{-1/4}$, respectively. The
curve endpoints reflect the completed numerical scan range for each
anisotropy.}
\label{fig:S-scaling-adiabatic}
\end{figure*}

Eigenmodes were followed by increasing $N$ and comparing the same mode at
successive resolutions. At the lowest resolution we computed the full
spectrum, retained finite growing modes in the prescribed real and imaginary
parts of $\sigma$, and then tracked the selected mode as the resolution was
raised. When the change between two successive resolutions became small
enough, the calculation switched to a shift-invert solve using the previous
eigenvalue and interpolated eigenvector as the initial guess. A mode was
accepted as converged when
\begin{equation}
    |\Delta\sigma|
    \leq
    {\tt atol}
    +
    {\tt rtol}\,\max\left(|\sigma_N|,|\sigma_{N-\Delta N}|\right),
\end{equation}
with ${\tt atol}=10^{-10}$ and ${\tt rtol}=10^{-5}$ in the production runs.
All eigenvalues plotted in
Figures~\ref{fig:dispersion-branches}--\ref{fig:dbeta-normalized} satisfied
this convergence criterion; the final accepted resolutions over the plotted
points span $N=160$--$2656$.
Modes that failed this resolution test before reaching the maximum allowed
resolution, or for which no eigenvalue with $\gamma>0$ remained after
filtering, were discarded. The selected tearing eigenvalues plotted
in Figures~\ref{fig:dispersion-branches}--\ref{fig:dbeta-normalized} were
non-oscillatory to the numerical tolerance of the mode-tracking procedure;
candidate roots with a significant imaginary part of $\sigma$ were not
included in these tearing-branch scans. The inner-layer thickness reported
below was measured from the converged eigenfunction on the $z>0$ side as the
first point where the magnitudes of the ideal and resistive terms in the
$z$-induction equation are equal,
$|ikB_0\delta v_z|=|\eta(\delta B_z''-k^2\delta B_z)|$, using interpolation
between adjacent collocation points where this ratio crosses unity.

With this setup, before using the asymptotic branches to estimate the
fastest-growing mode, we
compare them directly with numerical dispersion relations. Figure~\ref{fig:dispersion-branches}
shows the growth rate as a function of $\alpha=ka$ for $S=10^5$,
$Pr_m=0$, and $\beta_0=2$, for the three closures used in this work. The
symbols are growth rates computed with PSECAS, while the solid lines show the
asymptotic scalings derived above. The Coppi branch is evaluated from
Eq.~(\ref{eq:coppi_growth}) and is independent of $\Delta'$ at leading
order. The FKR branch is evaluated from Eq.~(\ref{eq:fkr_growth}) using the
fitted outer-region matching parameter $\Delta'_{\rm fit}$ given in
Eq.~(\ref{eq:Delta_empirical_fit_alpha}); the latter preserves the exact
marginal boundary $\alpha_c=\sqrt{\mathcal{A}/\mathcal{R}_0}$.
The outer-region input used in $\Delta'_{\rm fit}$ is checked directly
in Appendix~\ref{app:outer-stability-boundary} by comparing the numerically
integrated curvature coefficient $a_2/a_0$ with the compact theoretical
estimate and with the fitted interpolation.
Thus the Coppi curves are direct leading-order asymptotic predictions, whereas
the FKR curves combine the inner-layer scaling with a fitted finite-wavelength
approximation for $\Delta'$ constructed from $a_2/a_0$. This approximation
closely reproduces the numerical FKR branches for all three closures.

The comparison in Figure~\ref{fig:dispersion-branches} supports the
term balances within the tested parameter range, but the FKR part should be
read as consistency with the fitted outer interpolation rather than as an
independent verification of the compact analytic $\Delta'$ approximation.
At small $\alpha$, the numerical points follow the nonconstant-$\psi$ Coppi scaling
$\gamma\tau_A\propto\mathcal{A}^{1/3}\alpha^{2/3}S^{-1/3}$. At larger
$\alpha$, where the constant-$\psi$ ordering applies and the modes approach
the marginal cutoff, the numerical dispersions are described by the FKR
scaling with $\Delta'_{\rm fit}$. Negative $\Delta\beta_0$ increases the
effective tension factor $\mathcal{A}$, produces larger growth rates, and
extends the unstable interval to larger $\alpha$. Positive
$\Delta\beta_0$ has the opposite effect, suppressing the growth rate and
moving the stability cutoff to smaller wavenumber. The same branch
structure is seen for all three closures. The double-isothermal case shows the
largest displacement of the curves because its pressure response coefficient
vanishes and the imposed equilibrium anisotropy controls the dispersion
directly through $\mathcal{A}$ and $\mathcal{R}_0$.

\subsection{Parameter Dependence of the Fastest-Growing Mode}
\label{ssec:max-growth-parameter-scans}

We now compare the analytical trends derived above with numerical solutions
of the linear problem. The results are summarized in terms of the maximum
growth rate, $\gamma_{\max}\tau_A$, the corresponding wavenumber,
$k_{\max}a$, and the inner-layer thickness, $\delta_{\rm in}/a$. The scans
are performed in the inviscid limit, $Pr_m=0$. Throughout this section, $S$
is the Lundquist number based on the current-sheet thickness $a$, as defined
in Section~\ref{ssec:inner-scalings}.

As a first check on the resistive exponents in the fastest-mode
scalings, Figure~\ref{fig:S-scaling-adiabatic} shows an $S$ scan for the
double-adiabatic closure at $\beta_0=2$ and $Pr_m=0$.
The low-$S$ points are shown for context, but the
power-law fits use only the asymptotic part of the scan, $S\geq10^4$. For
$\Delta\beta_0=-0.50$, the fitted exponents are
$\gamma_{\max}\tau_A\propto S^{-0.49\pm0.06}$,
$k_{\max}a\propto S^{-0.23\pm0.06}$, and
$\delta_{\rm in}/a\propto S^{-0.25\pm0.06}$. For
$\Delta\beta_0=0$, the corresponding exponents are
$-0.49\pm0.06$, $-0.23\pm0.06$, and $-0.26\pm0.06$, and for
$\Delta\beta_0=+0.50$ they are $-0.49\pm0.06$, $-0.22\pm0.06$, and
$-0.26\pm0.06$. These measured slopes are consistent, within the fitted
uncertainties, with the classical dependences
$\gamma_{\max}\tau_A\propto S^{-1/2}$, $k_{\max}a\propto S^{-1/4}$, and
$\delta_{\rm in}/a\propto S^{-1/4}$.
The equilibrium anisotropy therefore changes the
normalization of the fastest-growing mode through $\mathcal{A}$ and
$\mathcal{R}_0$, but it does not alter the resistive $S$ exponents over the
large-$S$ part of the scan.
Since the closure parameters enter these fastest-mode estimates only through
the $S$-independent factor $\mathcal{R}_0$, changing the gyrotropic closure
modifies the prefactors but leaves the powers of $S$ unchanged.

\begin{figure*}[!ht]
\centering
\includegraphics[width=0.95\textwidth]{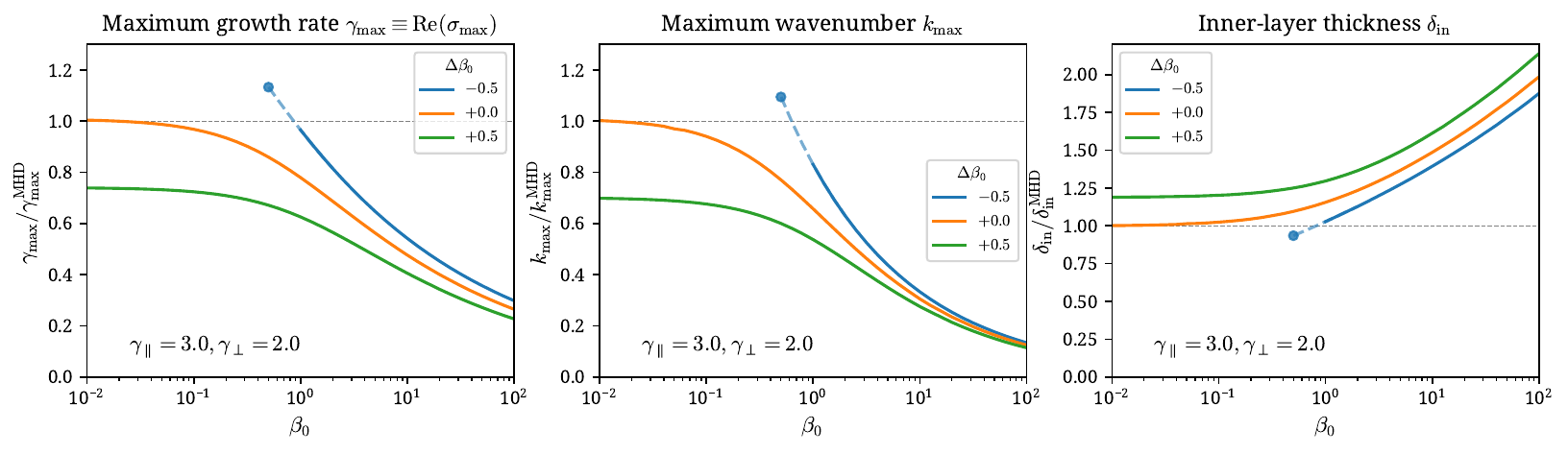}\\
\includegraphics[width=0.95\textwidth]{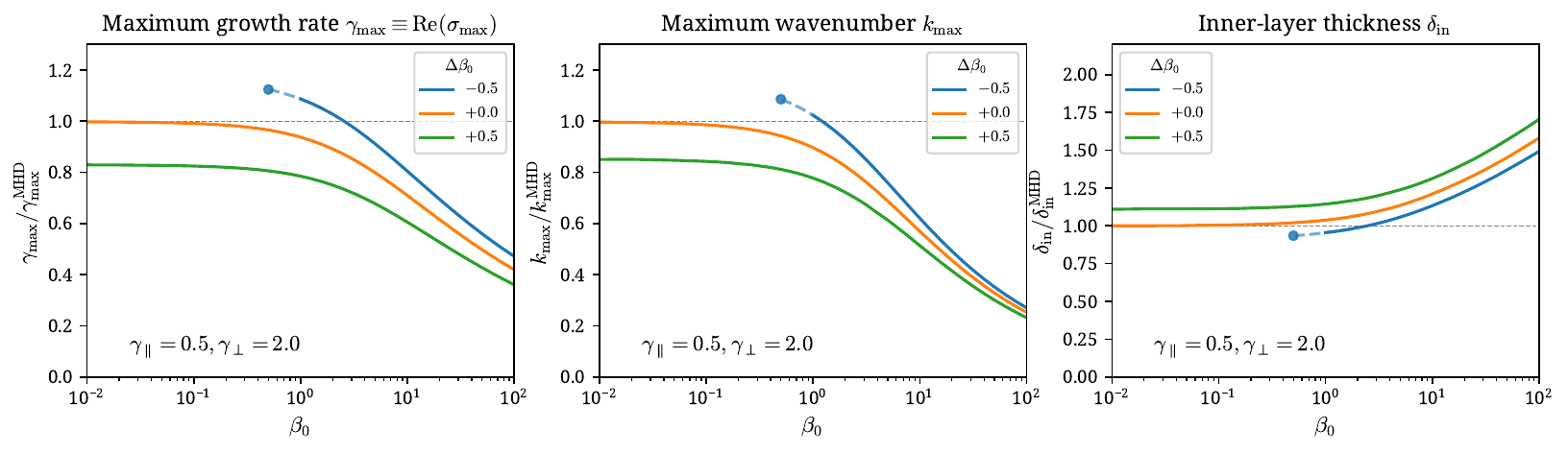}\\
\includegraphics[width=0.95\textwidth]{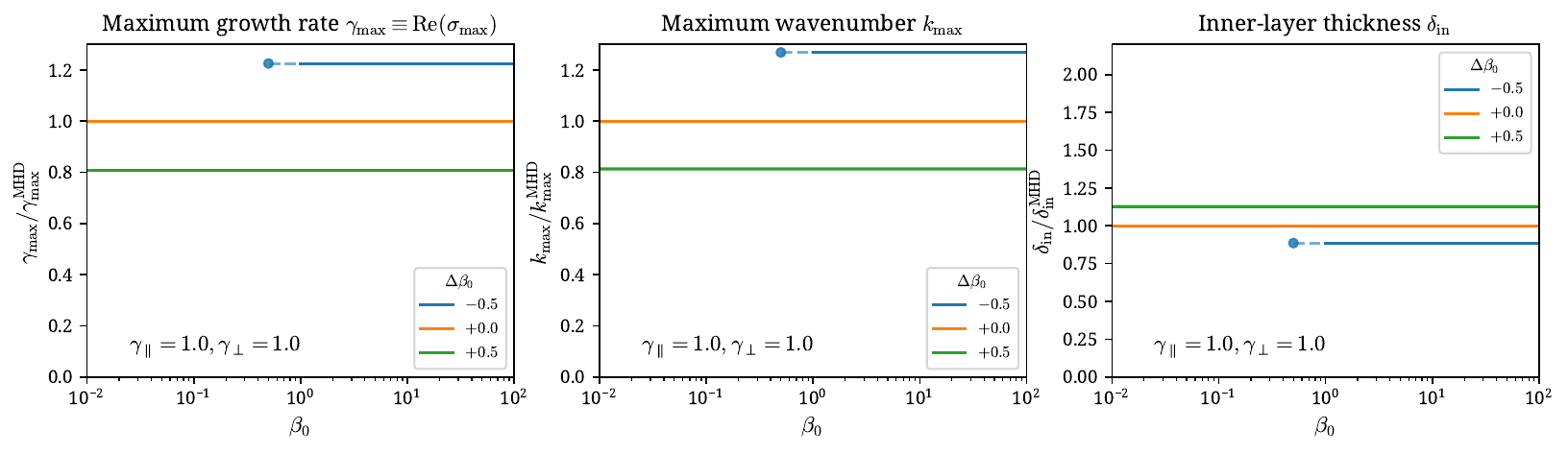}
\caption{Normalized dependence of the maximum tearing growth
rate, fastest-growing wavenumber, and inner-layer thickness on the equilibrium
plasma-$\beta$ for $S=10^5$ and $Pr_m=0$. Each quantity is normalized by the
corresponding classical MHD value at the same Lundquist number, indicated by
the horizontal dashed line. The three rows correspond, from top to bottom, to
the double-adiabatic closure $(\gamma_\parallel,\gamma_\perp)=(3,2)$, the
double-polytropic closure $(\gamma_\parallel,\gamma_\perp)=(0.5,2)$, and the
double-isothermal closure $(\gamma_\parallel,\gamma_\perp)=(1,1)$. In each
panel, the curves show $\Delta\beta_0=-0.5$, $0$, and $+0.5$. For the
$\Delta\beta_0=-0.5$ curve, semitransparent dashed segments mark the
mirror-unstable interval $\beta_0<1$, while the circular markers identify
the pressure-positivity boundary $p_{\parallel,0}=0$, i.e.
$\beta_0=0.5$. Missing curve segments correspond to values omitted by the
pressure-positivity, localized-eigenmode, or convergence filters.}
\label{fig:beta-normalized}
\end{figure*}

\begin{figure*}[!ht]
\centering
\includegraphics[width=0.95\textwidth]{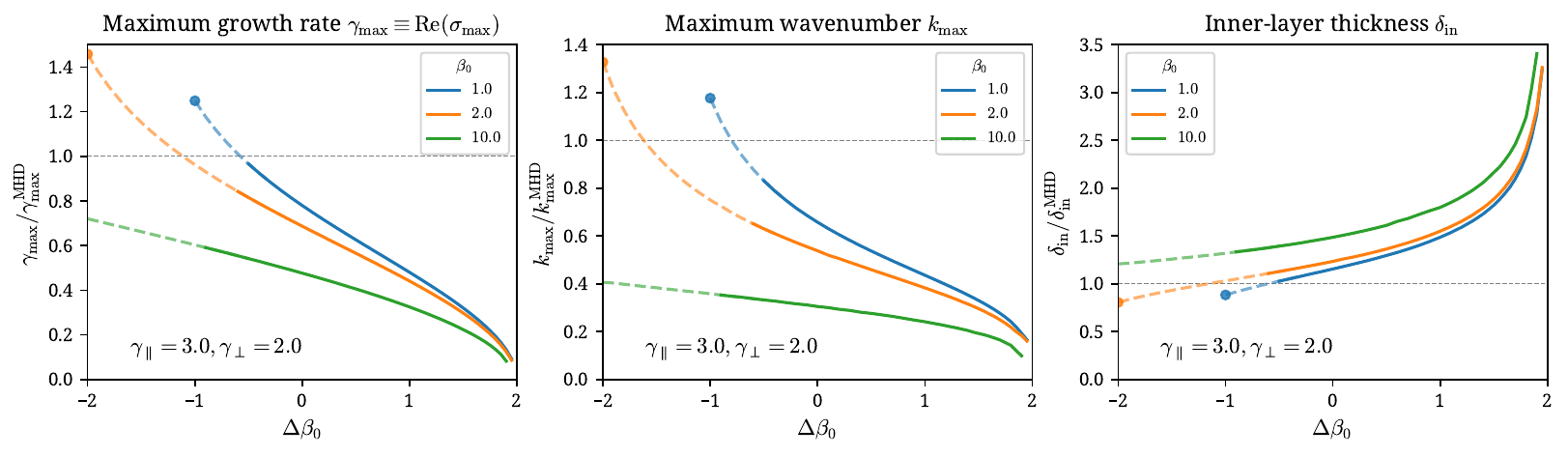}\\
\includegraphics[width=0.95\textwidth]{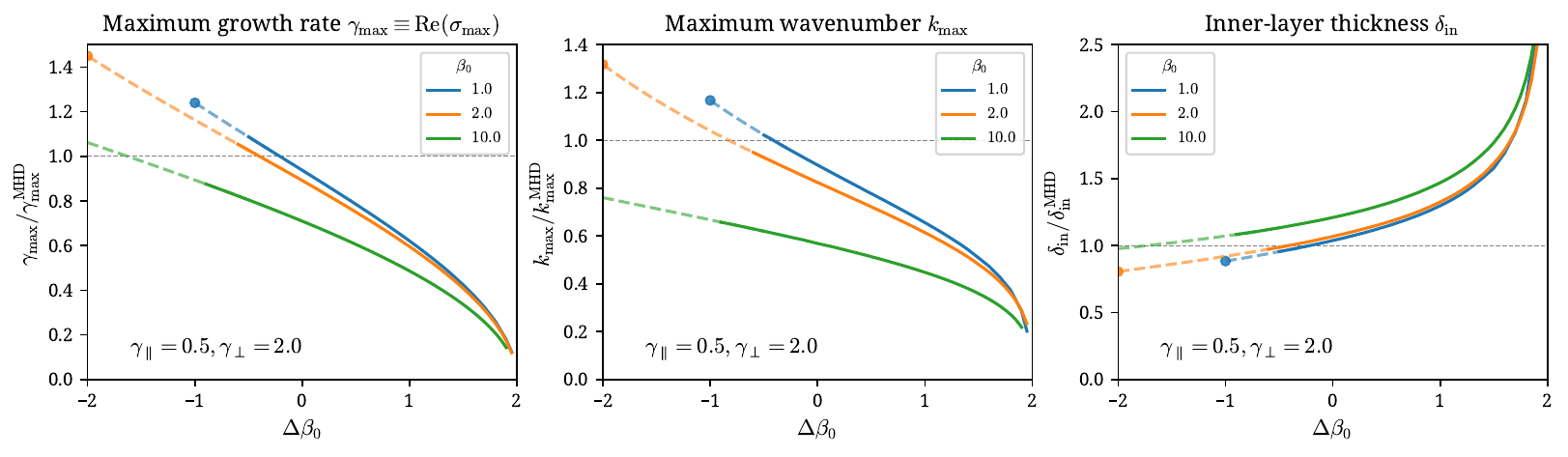}\\
\includegraphics[width=0.95\textwidth]{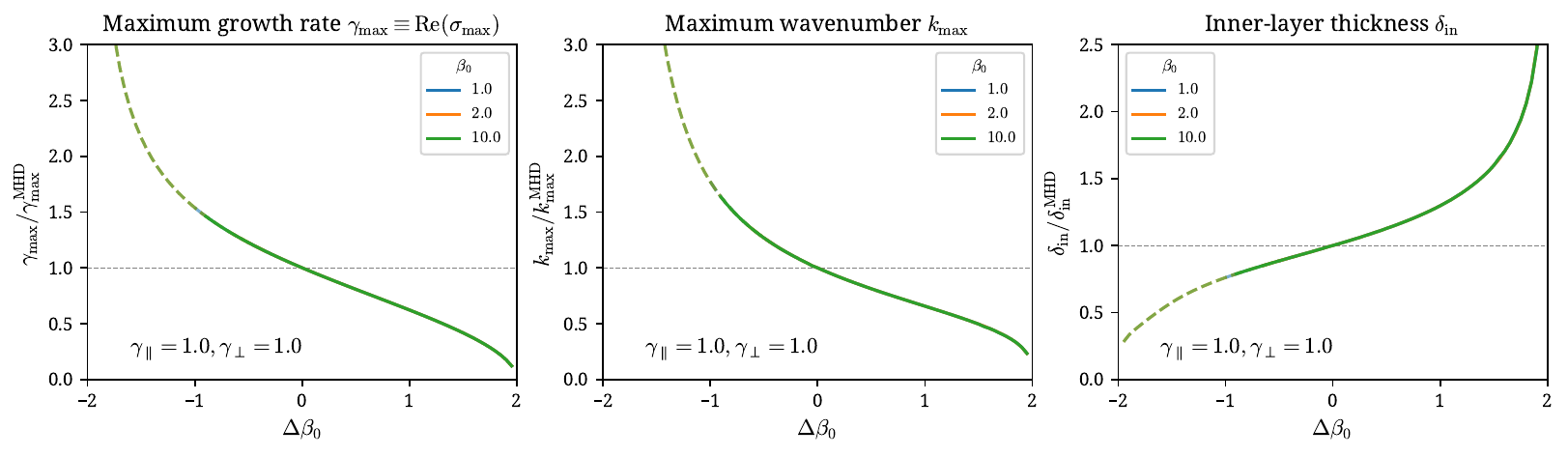}
\caption{Normalized dependence of the maximum tearing growth
rate, fastest-growing wavenumber, and inner-layer thickness on the prescribed
equilibrium anisotropy $\Delta\beta_0$ for $S=10^5$ and $Pr_m=0$. Each
quantity is normalized by the corresponding classical MHD value at the same
Lundquist number, indicated by the horizontal dashed line. The rows
correspond, from top to bottom, to the double-adiabatic closure
$(\gamma_\parallel,\gamma_\perp)=(3,2)$, the double-polytropic closure
$(\gamma_\parallel,\gamma_\perp)=(0.5,2)$, and the double-isothermal closure
$(\gamma_\parallel,\gamma_\perp)=(1,1)$. In each panel, the curves show
$\beta_0=1$, $2$, and $10$. Semitransparent dashed curve segments mark the
standard mirror-unstable interval
$\Delta\beta_0<-\beta_0/(1+\beta_0)$ for each color-coded $\beta_0$, while
the circular markers identify the pressure-positivity boundary
$p_{\parallel,0}=0$, or $\Delta\beta_0=-\beta_0$, where this boundary lies
inside the plotted interval. Missing curve segments
indicate values excluded by pressure positivity, loss of a localized tearing
eigenmode, or the convergence filter; the scans terminate as the
firehose/localization limit near $\Delta\beta_0=2$ is approached.}
\label{fig:dbeta-normalized}
\end{figure*}

We first consider the dependence on the equilibrium plasma-$\beta$. In this
scan, shown in Figure~\ref{fig:beta-normalized}, we fix $S=10^5$ and
normalize each quantity by its corresponding classical MHD value at the same
Lundquist number. This normalization removes the dominant resistive scaling
and isolates the gyrotropic correction. Each row corresponds to a different
closure, while the three curves in each panel correspond to
$\Delta\beta_0=-0.5$, $0$, and $+0.5$.
Physical admissibility in
Figures~\ref{fig:beta-normalized} and \ref{fig:dbeta-normalized} is indicated
directly in the scanned parameter space, rather than by mapping an exclusion
boundary through the asymptotic maximum-mode prefactors. We require
$p_{\parallel,0}>0$, equivalently $\Delta\beta_0>-\beta_0$, and use the
standard firehose/localization threshold $\Delta\beta_0=2$ and homogeneous
mirror threshold $\Delta\beta_0=-\beta_0/(1+\beta_0)$ as admissibility
guides. In Figure~\ref{fig:beta-normalized}, only the
$\Delta\beta_0=-0.5$ curve crosses these negative-anisotropy bounds in the
plotted $\beta_0$ range: the circular marker identifies
$p_{\parallel,0}=0$, or $\beta_0=0.5$, and the semitransparent dashed curve
segments mark the mirror-unstable interval $\beta_0<1$. In
Figure~\ref{fig:dbeta-normalized}, the same convention is applied at fixed
$\beta_0$: circular markers identify $\Delta\beta_0=-\beta_0$ when this
pressure-positivity boundary lies inside the plotted interval, and
semitransparent dashed curve segments mark
$\Delta\beta_0<-\beta_0/(1+\beta_0)$. These markers are not additional
closure-specific tearing eigenvalue boundaries; they only indicate where the
imposed equilibrium would be unstable to the standard anisotropy-driven
microinstabilities. A full oblique-mode microinstability analysis for each
gyrotropic closure is outside the scope of the present work.

For the double-adiabatic closure, the isotropic-equilibrium curve
($\Delta\beta_0=0$) reproduces the high-plasma-$\beta$ suppression found in the
previous isotropic-equilibrium analysis: the normalized maximum growth rate
and fastest-growing wavenumber decrease with increasing plasma-$\beta$, whereas
the inner layer becomes wider. A positive equilibrium anisotropy,
$\Delta\beta_0=+0.5$, further suppresses the instability and thickens the
inner layer. A negative anisotropy, $\Delta\beta_0=-0.5$, has the opposite
effect, producing larger growth rates and larger $k_{\max}$ over the range
where the localized tearing eigenmode exists. Nevertheless, even in this
case the plasma-$\beta$-dependent pressure-anisotropy response eventually drives the
mode toward smaller growth rates at sufficiently large plasma-$\beta$.

The double-polytropic closure shows the same qualitative behavior, but the
plasma-$\beta$ dependence is weaker. This is consistent with the smaller
coefficient multiplying the pressure-anisotropy fluctuations in the pressure
response. The double-isothermal closure behaves differently: for
$(\gamma_\parallel,\gamma_\perp)=(1,1)$, the plasma-$\beta$-dependent pressure response
vanishes, and the curves are essentially horizontal. In that case
the plasma-$\beta$ value $\beta_0$ does not change the normalized quantities, but the prescribed
equilibrium anisotropy still shifts the tearing mode relative to the MHD
reference value through its direct modification of the outer and inner
balances: negative $\Delta\beta_0$ enhances the maximum growth rate and
fastest-growing wavenumber while reducing the layer thickness, whereas
positive $\Delta\beta_0$ produces the opposite trend. The analytic reduction
of the double-isothermal outer problem is given in
Appendix~\ref{app:isothermal-outer}.

We next vary the prescribed equilibrium anisotropy $\Delta\beta_0$, as
shown in Figure~\ref{fig:dbeta-normalized}. The results use the same
normalization as in Figure~\ref{fig:beta-normalized}, so the horizontal
reference level corresponds to the classical resistive-MHD value. Increasing
$\Delta\beta_0$ corresponds to increasing the parallel pressure relative to
the perpendicular pressure.
The semitransparent dashed portions of the curves in
Figure~\ref{fig:dbeta-normalized} identify the formal continuation into the
standard mirror-unstable region,
$\Delta\beta_0<-\beta_0/(1+\beta_0)$, while the circular markers identify
the pressure-positivity boundary $\Delta\beta_0=-\beta_0$ for the
color-coded curves whose boundary lies inside the plotted interval.
Increasing $\Delta\beta_0$ also reduces the factor
$\mathcal{A}=1-\Delta\beta_0/2$, which controls the leading
anisotropic-tension contribution in the inner layer.
In the tearing balance this same term supplies the Alfv\'enic
accelerating response that couples the convective induction term to the
reconnecting field; reducing $\mathcal{A}$ therefore lowers the growth-rate
prefactor in Eqs.~(\ref{eq:fkr_growth}) and (\ref{eq:coppi_growth}) even
though it corresponds to weaker effective magnetic tension.

For the double-adiabatic and double-polytropic closures, the numerical
solutions show a systematic reduction of the maximum growth rate and
fastest-growing wavenumber as $\Delta\beta_0$ increases, together with a
broadening of the resistive layer. The effect is strongest as
$\Delta\beta_0$ approaches the upper localization boundary associated with
the loss of effective magnetic tension. Negative $\Delta\beta_0$, which
corresponds to $p_{\perp,0}>p_{\parallel,0}$, shifts the instability toward
larger growth rates, larger $k_{\max}$, and thinner inner layers. At fixed
$\Delta\beta_0$, increasing plasma-$\beta$ generally strengthens the stabilizing
pressure-anisotropy response for these two closures, so the $\beta_0=10$
curves lie at smaller normalized growth rates and wavenumbers and at larger
normalized layer thicknesses over most of the allowed interval.
Missing curve segments correspond to parameter values
outside the pressure-positive or localized-eigenmode domain, or to points
removed by the convergence filter.

The double-isothermal closure isolates the direct effect of the prescribed
equilibrium anisotropy, because the plasma-$\beta$-dependent pressure-fluctuation
response vanishes for
$(\gamma_\parallel,\gamma_\perp)=(1,1)$. Consequently, the curves for plasma-$\beta$ values
$\beta_0=1$, $2$, and $10$ nearly collapse. In this case negative
$\Delta\beta_0$ strongly enhances the tearing mode and moves the fastest
mode to much larger wavenumbers, while positive $\Delta\beta_0$ suppresses
the instability and broadens the inner layer as the firehose-related
tension boundary is approached. This behavior is consistent with the
analytical inner-layer balances, which predict that the imposed equilibrium
anisotropy modifies both the FKR and Coppi branches directly through
$\mathcal{A}$, in addition to its influence on the outer-region matching
parameter $\Delta'$.

Taken together, Figures~\ref{fig:beta-normalized} and
\ref{fig:dbeta-normalized} show that the equilibrium pressure anisotropy
changes the normalized prefactors and the location of the fastest-growing
mode in a closure-dependent way. These numerical trends are consistent with
the analytical term-balance result that $\Delta\beta_0$ enters both through
the outer-region matching parameter and directly through the inner-layer
factor $\mathcal{A}$.

\section{Discussion}
\label{sec:discussion}

\begin{table*}[!t]
\centering
\caption{Summary of leading gyrotropic corrections for the present force-free
current sheet. Here
$\mathcal{A}=1-\Delta\beta_0/2$ and
$\mathcal{R}_0=1+\frac{1}{2}(\gamma_\parallel+\gamma_\perp-2)\beta_0
+\frac{1}{2}\gamma_\parallel\Delta\beta_0$.
Dimensionless coefficients of order unity are omitted.}
\label{tab:gyro-substitutions}
\renewcommand{\arraystretch}{1.35}
\begin{tabular}{lll}
\hline
\hline
Quantity & Classical resistive MHD & Present gyrotropic-MHD result \\
\hline
Far-field decay rate
& $\lambda a=\alpha$
& $\lambda a=\alpha\sqrt{\mathcal{A}/\mathcal{R}_0}$ \\
Marginal cutoff
& $\alpha_c=1$
& $\alpha_c=\sqrt{\mathcal{A}/\mathcal{R}_0}$ \\
Long-wavelength tearing index
& $\Delta'\sim2/\alpha$
& $\Delta'\sim(2/\alpha)\sqrt{\mathcal{A}/\mathcal{R}_0}$ \\
FKR growth rate
& $\gamma\tau_A\sim\alpha^{2/5}(\Delta')^{4/5}S^{-3/5}$
& $\gamma\tau_A\sim\mathcal{A}^{1/5}\alpha^{2/5}
(\Delta')^{4/5}S^{-3/5}$, using gyrotropic $\Delta'$ \\
Coppi growth rate
& $\gamma\tau_A\sim\alpha^{2/3}S^{-1/3}$
& $\gamma\tau_A\sim\mathcal{A}^{1/3}\alpha^{2/3}S^{-1/3}$ \\
Fastest-growing wavenumber
& $\alpha_{\max}\sim S^{-1/4}$
& $\alpha_{\max}\sim\mathcal{A}^{1/4}\mathcal{R}_0^{-3/8}S^{-1/4}$ \\
Maximum growth rate
& $\gamma_{\max}\tau_A\sim S^{-1/2}$
& $\gamma_{\max}\tau_A\sim
\mathcal{A}^{1/2}\mathcal{R}_0^{-1/4}S^{-1/2}$ \\
Fastest-mode layer width
& $\delta_{\max}/a\sim S^{-1/4}$
& $\delta_{\max}/a\sim
\mathcal{A}^{-1/4}\mathcal{R}_0^{1/8}S^{-1/4}$ \\
\hline
\end{tabular}
\end{table*}

The analysis shows that a prescribed equilibrium pressure anisotropy modifies
resistive tearing through two distinct channels. The first is the ideal outer
region, where the anisotropy changes the coefficient multiplying the magnetic
perturbation and therefore changes both the far-field decay rate and the
outer matching parameter $\Delta'$. The second is the resistive inner layer,
where the same imposed anisotropy appears as the effective tension factor
$\mathcal{A}=1-\Delta\beta_0/2$ in the leading momentum balance. This
separation is useful because the Coppi branch is independent of $\Delta'$ at
leading order but still depends directly on $\mathcal{A}$, whereas the FKR
branch depends on both $\mathcal{A}$ and the outer solution.

The stability boundary is controlled by the global outer-region problem. The
exact marginal condition,
$\alpha_c^2=\mathcal{A}/\mathcal{R}_0$, follows from the outer equation and
is reproduced by the numerical integration of the outer region. This point is
important for the FKR branch because the growth rate must vanish at the same
cutoff that bounds the localized outer eigenfunction. The empirical
$\Delta'_{\rm fit}$ used in the branch comparison keeps this exact
boundary while improving the finite-wavelength behavior of the outer matching
parameter. The agreement of the fitted FKR branch with the numerical
dispersion relations in Figure~\ref{fig:dispersion-branches} therefore
supports both the asymptotic inner-layer balances and the global outer
stability condition.

The closure dependence is carried mainly by the pressure-anisotropy response
coefficient entering $\mathcal{R}_0$. For the double-adiabatic and
double-polytropic closures, increasing plasma-$\beta$ strengthens the
pressure-response correction and generally suppresses the fastest tearing
mode, especially for positive $\Delta\beta_0$. Negative
$\Delta\beta_0$, corresponding to $p_{\perp,0}>p_{\parallel,0}$, increases
$\mathcal{A}$ and enhances the instability over the parameter range where the
localized mode and the equilibrium pressures remain physical. The
double-isothermal closure is a useful limiting case because the
pressure-fluctuation response vanishes. In that limit the numerical scans are
essentially independent of plasma-$\beta$, so the changes in the dispersion
relation isolate the direct effect of the prescribed equilibrium anisotropy
through $\mathcal{A}$ and $\mathcal{R}_0$.

The maximum-growth scalings retain the classical resistive exponents,
$\gamma_{\max}\tau_A\propto S^{-1/2}$,
$k_{\max}a\propto S^{-1/4}$, and
$\delta_{\max}/a\propto S^{-1/4}$, but the prefactors are no longer universal.
For fixed closure parameters the leading estimates give
$\gamma_{\max}\tau_A\propto
\mathcal{A}^{1/2}\mathcal{R}_0^{-1/4}$ and
$k_{\max}a\propto\mathcal{A}^{1/4}\mathcal{R}_0^{-3/8}$. Thus equilibrium
anisotropy changes the location and strength of the fastest tearing mode
without changing the Lundquist-number exponents in the regime considered
here. These scalings should not be extrapolated across the boundaries where
either $\mathcal{A}$ or $\mathcal{R}_0$ ceases to be positive, or where
pressure positivity is lost, because the unstable nonoscillatory inner-layer
balance or the localized quasistatic outer solution then ceases to provide the
relevant matching problem.

For readers interested in applying these results, Table~\ref{tab:gyro-substitutions}
summarizes the leading changes that convert the classical resistive-MHD
estimates into the gyrotropic scalings derived here. The two required inputs
are the effective inner-layer tension factor $\mathcal{A}$ and the modified
global outer response, represented by $\mathcal{R}_0$ and the corresponding
gyrotropic tearing index $\Delta'$.

These replacements apply to the incompressible, inviscid, force-free
Harris-type equilibrium considered here and to the unstable nonoscillatory
localized branch with $\mathcal{A}>0$ and $\mathcal{R}_0>0$. For a different
current-sheet profile, the global outer solution, $\Delta'$, and the marginal
cutoff must generally be recalculated; the resulting fastest-mode prefactors
therefore cannot be inferred from the table without a new outer-region
matching calculation.

Several restrictions should be kept in mind. The equilibrium anisotropy is
prescribed rather than evolved self-consistently, and possible kinetic
relaxation by mirror, firehose, or heat-flux physics is not included. The
analysis is also incompressible, assumes the force-free Harris-type current
sheet used here, and focuses on the linear, inviscid tearing problem. These
assumptions make the outer--inner matching analytically transparent, but they
leave open how the same anisotropic corrections would be modified by density
perturbations, compressible pressure balance, nonlinear island evolution,
viscous effects, or microinstability-limited pressure anisotropy.
Thus the astrophysical relevance of the results is conditional: they
show how a prescribed equilibrium anisotropy enters a controlled
gyrotropic-MHD onset calculation, but they do not by themselves determine the
onset of reconnection in systems where the anisotropy is generated, bounded,
or relaxed by kinetic physics.

The frozen-equilibrium ordering introduces an additional restriction. The
prescribed force-free profile evolves resistively under the full nonideal
equations, so the eigenvalue calculation applies when the tearing growth time
is short compared with the equilibrium diffusion time, approximately
$\gamma\tau_A\gg S^{-1}$. Accordingly, the exact outer marginal boundary is a
property of the frozen-profile stability problem; the immediate neighborhood
of $\gamma=0$ should not be interpreted as the long-time evolution of an
unmaintained resistive current sheet.

Natural extensions of this work would relax these assumptions one at a time.
A compressible formulation would allow the tearing mode to couple to density
and acoustic responses, while equilibria with nonuniform parallel and
perpendicular thermal-pressure profiles could be constructed together with the
guide-field profile instead of being imposed on a force-free sheet with
uniform pressures. A fully three-dimensional treatment would also permit
oblique tearing modes and possible coupling to guide-field-dependent
instabilities that are absent in the two-dimensional geometry considered
here. Finally, when the inner layer approaches ion kinetic scales, Hall terms
and more general two-fluid effects should modify the induction equation and
the pressure response. Incorporating these ingredients would test how much of
the present gyrotropic-MHD scaling structure survives beyond the resistive,
single-fluid limit.

\section{Conclusions}
\label{sec:conclusions}

We have developed an analytical and numerical theory for the linear tearing
instability of a force-free current sheet in nonideal gyrotropic MHD with a
prescribed equilibrium pressure anisotropy. The calculation keeps the
classical outer--inner tearing structure but removes the assumption that the
equilibrium pressure tensor is isotropic. This makes it possible to isolate
how a finite
$\Delta\beta_0=\beta_{\parallel,0}-\beta_{\perp,0}$ modifies both the ideal
outer problem and the resistive inner-layer dynamics.

This extension is useful as a controlled model because the classical
incompressible MHD tearing scalings are insensitive to plasma-$\beta$, whereas
weakly collisional space and astrophysical plasmas can support different
parallel and perpendicular pressures. Our results show that, when a prescribed
equilibrium anisotropy is allowed within this reduced single-fluid framework,
it changes the linear tearing stability problem and the fastest-growing mode.
They should therefore be interpreted as conditional gyrotropic-MHD corrections
to the onset calculation, not as a standalone prediction for fully kinetic
astrophysical current sheets.

The main results are as follows.

\begin{enumerate}
\item Starting from the nonideal gyrotropic-MHD equations, we derived the
linear perturbation system for a frozen force-free Harris-type sheet with uniform
parallel and perpendicular equilibrium pressures. The equilibrium anisotropy
enters the analytical theory through the effective tension factor
$\mathcal{A}=1-\Delta\beta_0/2$ and through the pressure-response coefficient
$\mathcal{R}_0=1+\frac{1}{2}
\left[(\gamma_\parallel+\gamma_\perp-2)\beta_0
+\gamma_\parallel\Delta\beta_0\right]$.

\item In the ideal outer region, the prescribed anisotropy changes the
far-field decay rate and the tearing stability index $\Delta'$. The localized
frozen-profile marginal boundary is set by the global outer solution and is
\[
\alpha_c^2=\frac{\mathcal{A}}{\mathcal{R}_0},
\]
in agreement with direct numerical integration of the outer equation. The
exact zero-wavenumber outer solution gives the long-wavelength
coefficient $\mathcal{C}_\Delta=2\sqrt{\mathcal{A}/\mathcal{R}_0}$ in
$\Delta'\sim\mathcal{C}_\Delta/\alpha$. The compact approximation in
Eq.~(\ref{eq:Delta_global_closure}) preserves this coefficient and the exact
marginal root, while the fitted expression $\Delta'_{\rm fit}$ used in the
branch comparison improves the finite-wavelength behavior and preserves the
same exact marginal condition.

\item In the resistive inner layer, the imposed anisotropy enters the leading
momentum balance through $\mathcal{A}$. Therefore, the FKR branch depends on
anisotropy through both the outer matching parameter $\Delta'$ and the
inner-layer force balance, while the Coppi branch depends directly on
$\mathcal{A}$ even though it is independent of $\Delta'$ at leading order.

\item For the unstable nonoscillatory branch with
$\mathcal{A}>0$ and $\mathcal{R}_0>0$, matching the FKR and Coppi branches
gives, up to order-unity constants,
\[
\begin{aligned}
k_{\max}a
&\sim\mathcal{A}^{1/4}\mathcal{R}_0^{-3/8}S^{-1/4},\\
\gamma_{\max}\tau_A
&\sim\mathcal{A}^{1/2}\mathcal{R}_0^{-1/4}S^{-1/2},\\
\frac{\delta_{\max}}{a}
&\sim\mathcal{A}^{-1/4}\mathcal{R}_0^{1/8}S^{-1/4}.
\end{aligned}
\]
Thus the classical Lundquist-number exponents are retained in the regime
considered here, but the prefactors are no longer universal.

\item
Numerical dispersion relations computed with PSECAS are consistent with the
Coppi branch and with the FKR branch when a fitted finite-wavelength
approximation for $\Delta'$ is used, for the three closures considered in this
work:
double-adiabatic, double-polytropic, and double-isothermal. The fitted
$a_2/a_0$ form entering $\Delta'_{\rm fit}$ allows the calculated FKR curves
to closely follow the numerical branches for all three closures. The branch
comparison supports the conclusion that the anisotropic modification of the
outer solution and the direct inner-layer tension correction are both required
to describe the numerical eigenvalues in the tested parameter range.

\item The numerical scans show a closure-dependent modification of the
maximum growth rate, fastest-growing wavenumber, and inner-layer thickness.
Positive $\Delta\beta_0$ generally suppresses tearing within the
localized-mode, pressure-positive domain, narrows the unstable
wavenumber interval, and broadens the inner layer, whereas negative
$\Delta\beta_0$ enhances the instability over the allowed parameter range. In
the double-isothermal limit the explicit plasma-$\beta$ dependence
disappears, leaving only the direct effect of the imposed equilibrium
anisotropy through $\mathcal{A}$ and $\mathcal{R}_0$.
\end{enumerate}

The main consequence of removing the pressure-isotropy assumption is that the
tearing problem is no longer governed only by the magnetic geometry and the
resistive scale separation. A prescribed equilibrium pressure anisotropy
changes the ideal matching problem, the far-field localization condition, the
marginal stability boundary, and the leading inner-layer force balance. These
effects survive even in the simplified force-free configuration used here,
where equilibrium pressure-gradient forces have been intentionally removed.

Within this controlled model, an isotropic-MHD description can miss
order-unity changes in the tearing growth rate, preferred wavelength, and
inner-layer thickness when an equilibrium pressure anisotropy is imposed. This
motivates extending tearing calculations beyond pressure isotropy for
high-$\beta$ or weakly collisional environments, while keeping in mind that
quantitative applications require the additional physics listed above. The
present gyrotropic-MHD calculation provides one controlled step in that
direction and identifies where extensions such as compressibility,
self-consistent pressure profiles, Hall physics, and two-fluid effects should
enter.

\begin{acknowledgments}
The authors acknowledge support from FAPESP (grants 2013/10559-5,
2021/02120-0, 2021/06502-4, 2022/03972-2, and 2024/16327-3). The numerical
eigenvalue runs with the PSECAS solver presented in this work were carried out
on the Hydra HPC cluster of the Group of Theoretical Astrophysics at EACH-USP,
which was acquired with support from FAPESP (grants 2013/04073-2 and
2022/03972-2). The authors used OpenAI Codex \citep{OpenAICodex2026} to assist
with language editing, manuscript organization, and checks of algebraic
derivations. All authors reviewed the text, calculations, citations, and
conclusions and take full responsibility for the manuscript.
\end{acknowledgments}

\clearpage
\onecolumngrid
\restartappendixnumbering
\appendix
\renewcommand{\theequation}{\thesection\arabic{equation}}
\renewcommand{\theHequation}{\thesection\arabic{equation}}

\section{Useful Relations for the Equilibrium Magnetic Field}
\label{app:equilibrium-identities}
\setcounter{equation}{0}

The equilibrium magnetic field used throughout this work is
\begin{align}
    \mathbf{B}_0(z)
    &=B_0(z)\hat{\mathbf{i}}+B_g(z)\hat{\mathbf{j}}, \\
    B_0(z)&=\tanh\left(\frac{z}{a}\right),
    &
    B_g(z)&=\sech\left(\frac{z}{a}\right),
\end{align}
so that
\begin{equation}
    B_0^2+B_g^2=1.
\end{equation}
Here primes denote derivatives with respect to $z$. The first derivatives are
\begin{align}
    B_0' &= \frac{1}{a}\sech^2\left(\frac{z}{a}\right)
          = \frac{1}{a}B_g^2
          = \frac{1}{a}\left(1-B_0^2\right), \\
    B_g' &= -\frac{1}{a}\sech\left(\frac{z}{a}\right)
          \tanh\left(\frac{z}{a}\right) \nonumber\\
          = -\frac{1}{a}B_0 B_g .
\end{align}
The second derivatives can therefore be written as
\begin{align}
    B_0'' &= -\frac{2}{a^2}B_0 B_g^2 \nonumber\\
           &= -\frac{2}{a^2}B_0\left(1-B_0^2\right), \\
    B_g'' &= \frac{1}{a^2}B_g\left(B_0^2-B_g^2\right) \nonumber\\
           &= \frac{1}{a^2}B_g\left(2B_0^2-1\right) \nonumber\\
           &= \frac{1}{a^2}B_g\left(1-2B_g^2\right).
\end{align}
Differentiating $B_0^2+B_g^2=1$ also gives the identities
\begin{align}
    B_0 B_0' + B_g B_g' &= 0, \\
    B_g B_0' - B_0 B_g' &= \frac{1}{a}B_g, \\
    B_0 B_0'' + B_g B_g'' &= -\left(B_0'^2+B_g'^2\right)
                           = -\frac{1}{a^2}B_g^2 .
\end{align}
The combinations that enter the linearized equations are
\begin{align}
    B_gB_0'-B_0B_g' &= \frac{B_g}{a}, \\
    B_0B_g'+2B_0'B_g
    &= \frac{B_g}{a}\left(2B_g^2-B_0^2\right).
\end{align}
These identities retain the current-sheet thickness explicitly.

\section{Derivation of the Linearized Gyrotropic MHD Equations}
\label{app:linearization}
\setcounter{equation}{0}

This appendix starts from the compressible nonideal gyrotropic equations and
then specializes them to the incompressible, 2.5D geometry used in the main
text. We use units in which $\mu_0=1$, write
$d/dt=\partial_t+\mathbf{v}\cdot\nabla$, and assume constant $\eta$ and
$\nu$.

\subsection{Compressible System}

The compressible model is
\begin{align}
    \frac{d\rho}{dt} &= -\rho\nabla\cdot\mathbf{v}, \\
    \rho\frac{d\mathbf{v}}{dt}
    &= -\nabla\cdot\mathsf{P}
    +\mathbf{J}\times\mathbf{B}
    +\rho\nu\left[\nabla^2\mathbf{v}
    +\frac{1}{3}\nabla(\nabla\cdot\mathbf{v})\right], \\
    \frac{\partial\mathbf{B}}{\partial t}
    &= \nabla\times(\mathbf{v}\times\mathbf{B}-\eta\mathbf{J}),
    \qquad
    \mathbf{J}=\nabla\times\mathbf{B},
    \qquad
    \nabla\cdot\mathbf{B}=0,
\end{align}
with pressure tensor
\begin{equation}
    \mathsf{P}=p_\perp\mathsf{I}+\Delta p\,\mathbf{b}\mathbf{b},
    \qquad
    \Delta p=p_\parallel-p_\perp,
    \qquad
    \mathbf{b}=\frac{\mathbf{B}}{B}.
\end{equation}
Using $\nabla\cdot\mathbf{B}=0$, the induction equation may also be written
as
\begin{equation}
    \frac{\partial\mathbf{B}}{\partial t}
    =
    (\mathbf{B}\cdot\nabla)\mathbf{v}
    -(\mathbf{v}\cdot\nabla)\mathbf{B}
    -\mathbf{B}(\nabla\cdot\mathbf{v})
    +\eta\nabla^2\mathbf{B}.
\end{equation}
The pressure equations are
\begin{align}
    \frac{dp_\parallel}{dt}
    &=
    -p_\parallel\nabla\cdot\mathbf{v}
    -(\gamma_\parallel-1)p_\parallel
    \mathbf{b}\cdot(\mathbf{b}\cdot\nabla)\mathbf{v}
    +(\gamma_\parallel-1)\left[
    \eta(\mathbf{b}\cdot\mathbf{J})^2+\frac{1}{3}Q_\nu
    \right], \\
    \frac{dp_\perp}{dt}
    &=
    -\gamma_\perp p_\perp\nabla\cdot\mathbf{v}
    +(\gamma_\perp-1)p_\perp
    \mathbf{b}\cdot(\mathbf{b}\cdot\nabla)\mathbf{v}
    +(\gamma_\perp-1)\left[
    \eta\left\{\mathbf{J}\cdot\mathbf{J}
    -(\mathbf{b}\cdot\mathbf{J})^2\right\}
    +\frac{2}{3}Q_\nu
    \right],
\end{align}
where
\begin{equation}
    Q_\nu\equiv
    \rho\nu\left[
    (\nabla\times\mathbf{v})^2
    -\frac{4}{3}\mathbf{v}\cdot\nabla(\nabla\cdot\mathbf{v})
    \right].
\end{equation}
Subtracting the two pressure equations gives
\begin{equation}
\begin{split}
    \frac{d\Delta p}{dt}
    &=
    -(p_\parallel-\gamma_\perp p_\perp)\nabla\cdot\mathbf{v}
    -\left[(\gamma_\parallel-1)p_\parallel
    +(\gamma_\perp-1)p_\perp\right]
    \mathbf{b}\cdot(\mathbf{b}\cdot\nabla)\mathbf{v} \\
    &\quad
    +\eta\left[
    (\gamma_\parallel+\gamma_\perp-2)(\mathbf{b}\cdot\mathbf{J})^2
    -(\gamma_\perp-1)(\mathbf{J}\cdot\mathbf{J})
    \right]
    +\frac{1}{3}(\gamma_\parallel-2\gamma_\perp+1)Q_\nu .
\end{split}
\label{eq:app_compressible_deltap}
\end{equation}

\subsection{Incompressible Reduction}

For the incompressible problem, $\nabla\cdot\mathbf{v}=0$ and
$\rho=1$. The velocity equation becomes Eq.~(\ref{eq:momentum}), while the
induction equation becomes Eq.~(\ref{eq:induction}). The pressure-difference
equation reduces to
\begin{equation}
\begin{split}
    \frac{\partial\Delta p}{\partial t}
    +\mathbf{v}\cdot\nabla\Delta p
    &=
    -C_p\,\mathbf{b}\cdot(\mathbf{b}\cdot\nabla)\mathbf{v}
    +\eta\left[
    G(\mathbf{b}\cdot\mathbf{J})^2
    -H(\mathbf{J}\cdot\mathbf{J})
    \right]
    +\frac{1}{3}(\gamma_\parallel-2\gamma_\perp+1)\nu\omega^2,
\end{split}
\label{eq:app_incompressible_deltap}
\end{equation}
where
\begin{align}
    C_p&=(\gamma_\parallel+\gamma_\perp-2)p_\perp
    +(\gamma_\parallel-1)\Delta p, \\
    G&=\gamma_\parallel+\gamma_\perp-2,
    \qquad
    H=\gamma_\perp-1,
    \qquad
    \boldsymbol{\omega}=\nabla\times\mathbf{v}.
\end{align}
Taking the curl of Eq.~(\ref{eq:momentum}) removes the scalar
$p_\perp$ contribution. Since $\nabla\cdot\mathbf{B}=0$,
\begin{equation}
    \nabla\cdot(\Delta p\,\mathbf{b}\mathbf{b})
    =
    \mathbf{B}\cdot\nabla\left(\frac{\Delta p}{B^2}\mathbf{B}\right).
\end{equation}
Using also $\nabla\cdot\mathbf{J}=0$, the incompressible vorticity equation
is
\begin{equation}
    \frac{\partial\boldsymbol{\omega}}{\partial t}
    =
    -(\mathbf{v}\cdot\nabla)\boldsymbol{\omega}
    +(\boldsymbol{\omega}\cdot\nabla)\mathbf{v}
    +(\mathbf{B}\cdot\nabla)\mathbf{J}
    -(\mathbf{J}\cdot\nabla)\mathbf{B}
    -\nabla\times\left[
    \mathbf{B}\cdot\nabla\left(\frac{\Delta p}{B^2}\mathbf{B}\right)
    \right]
    +\nu\nabla^2\boldsymbol{\omega}.
    \label{eq:app_vorticity}
\end{equation}

\subsection{Equilibrium, Perturbations, and Constraints}

The equilibrium used in the paper is
\begin{equation}
    \mathbf{v}_0=0,
    \qquad
    \mathbf{B}_0=B_0(z)\hat{\mathbf{i}}+B_g(z)\hat{\mathbf{j}},
\end{equation}
with
\begin{align}
    B_0(z)&=\tanh\left(\frac{z}{a}\right),
    &
    B_g(z)&=\sech\left(\frac{z}{a}\right), \\
    B_0^2+B_g^2&=1.
\end{align}
The pressure components are uniform:
\begin{equation}
    p_{\perp,0}=\frac{\beta_0}{2},
    \qquad
    p_{\parallel,0}=\frac{\beta_0+\Delta\beta_0}{2},
    \qquad
    \Delta p_0=\frac{\Delta\beta_0}{2}.
\end{equation}
As stated in Section~\ref{sec:equations}, this profile is treated as a frozen
equilibrium: its zeroth-order nonideal evolution is omitted or externally
balanced, while first-order resistive contributions are retained in the
linearized perturbation equations.
Perturbations are taken as
\begin{equation}
    f(t,x,z)=f_0(z)+\delta f(z)e^{ikx+\sigma t},
    \qquad
    \partial_y=0,
\end{equation}
and primes denote $d/dz$. It is useful to define
\begin{equation}
    \mathcal{L}\equiv \frac{d^2}{dz^2}-k^2 .
\end{equation}
Here, as in the main text, $\sigma$ denotes the complex eigenvalue and
$\gamma={\rm Re}(\sigma)$ is the growth rate.
The incompressibility and solenoidal constraints give
\begin{equation}
    \delta v_x=\frac{i}{k}\delta v_z',
    \qquad
    \delta B_x=\frac{i}{k}\delta B_z' .
    \label{eq:app_constraints}
\end{equation}
The equilibrium and perturbed currents are
\begin{align}
    \mathbf{J}_0
    &= -B_g'\hat{\mathbf{i}}+B_0'\hat{\mathbf{j}}, \\
    \delta\mathbf{J}
    &= -\delta B_y'\hat{\mathbf{i}}
    +\left(\delta B_x'-ik\delta B_z\right)\hat{\mathbf{j}}
    +ik\delta B_y\hat{\mathbf{k}} .
    \label{eq:app_currents}
\end{align}

\subsection{Linearized Induction Equation}

The linearized induction equation is
\begin{equation}
    \sigma\delta\mathbf{B}
    =
    (\mathbf{B}_0\cdot\nabla)\delta\mathbf{v}
    -(\delta\mathbf{v}\cdot\nabla)\mathbf{B}_0
    +\eta\nabla^2\delta\mathbf{B}.
\end{equation}
In Fourier form its components are
\begin{align}
    \sigma\delta B_x
    &=ikB_0\delta v_x-B_0'\delta v_z+\eta\mathcal{L}\delta B_x, \\
    \sigma\delta B_y
    &=ikB_0\delta v_y-B_g'\delta v_z+\eta\mathcal{L}\delta B_y, \\
    \sigma\delta B_z
    &=ikB_0\delta v_z+\eta\mathcal{L}\delta B_z.
    \label{eq:app_induction_z}
\end{align}
Equations~(\ref{dby}) and (\ref{dbz}) are the last two equations above; the
$x$ component is redundant once Eq.~(\ref{eq:app_constraints}) is imposed.

\subsection{Linearized Vorticity Equation}

Because $\mathbf{v}_0=\boldsymbol{\omega}_0=0$, the nonlinear velocity terms
in Eq.~(\ref{eq:app_vorticity}) do not contribute at first order. The
linearized vorticity equation is therefore
\begin{equation}
    \sigma\delta\boldsymbol{\omega}
    =\mathbf{C}_{\rm L}-\mathbf{C}_{\Delta}
    +\nu\mathcal{L}\delta\boldsymbol{\omega},
    \label{eq:app_lin_vort}
\end{equation}
where
\begin{align}
    \mathbf{C}_{\rm L}
    &=(\mathbf{B}_0\cdot\nabla)\delta\mathbf{J}
    +(\delta\mathbf{B}\cdot\nabla)\mathbf{J}_0 \nonumber\\
    &\quad
    -(\mathbf{J}_0\cdot\nabla)\delta\mathbf{B}
    -(\delta\mathbf{J}\cdot\nabla)\mathbf{B}_0, \\
    \mathbf{C}_{\Delta}
    &=\nabla\times\delta\mathbf{F}_{\Delta}.
\end{align}
Here
$\mathbf{F}_{\Delta}\equiv
\mathbf{B}\cdot\nabla[(\Delta p/B^2)\mathbf{B}]$.
The linear perturbation of this anisotropic-force vector is
\begin{equation}
\begin{split}
    \delta\left[
    \mathbf{B}\cdot\nabla\left(\frac{\Delta p}{B^2}\mathbf{B}\right)
    \right]
    &=
    (\mathbf{B}_0\cdot\nabla)(\mathbf{B}_0\delta\Delta p) \\
    &\quad
    +\Delta p_0\left[
    (\mathbf{B}_0\cdot\nabla)\delta\mathbf{B}
    +(\delta\mathbf{B}\cdot\nabla)\mathbf{B}_0
    -2(\mathbf{B}_0\cdot\nabla)
    \left\{\mathbf{B}_0(\mathbf{B}_0\cdot\delta\mathbf{B})\right\}
    \right].
\end{split}
\label{eq:app_anis_force}
\end{equation}
The components needed for the independent velocity equations are
\begin{align}
    (\mathbf{C}_{\rm L})_y
    &=-B_0\mathcal{L}\delta B_z+B_0''\delta B_z, \\
    (\mathbf{C}_{\rm L})_z
    &=-k^2B_0\delta B_y+ikB_g'\delta B_z,
    \label{eq:app_lorentz_components}
\end{align}
and
\begin{align}
    (\mathbf{C}_{\Delta})_y
    &=ikB_0\left(2B_0'\delta\Delta p+B_0\delta\Delta p'\right) \nonumber\\
    &\quad
    +\Delta p_0\Bigl[
    (2B_0^3-B_0)\delta B_z''
    +6B_0^2B_0'\delta B_z'
    +B_0''\delta B_z
    +k^2B_0\delta B_z \nonumber\\
    &\hspace{1.15in}
    -2ikB_0^2B_g\delta B_y'
    -2ikB_0\left(B_0B_g'+2B_0'B_g\right)\delta B_y
    \Bigr], \\
    (\mathbf{C}_{\Delta})_z
    &=-k^2B_0B_g\delta\Delta p \nonumber\\
    &\quad
    +\Delta p_0\left[
    2ikB_0^2B_g\delta B_z'
    -k^2B_0(1-2B_g^2)\delta B_y
    +ikB_g'\delta B_z
    \right].
    \label{eq:app_anis_components}
\end{align}
Finally,
\begin{equation}
    \delta\omega_y=\frac{i}{k}\mathcal{L}\delta v_z,
    \qquad
    \delta\omega_z=ik\delta v_y .
\end{equation}
Substituting Eqs.~(\ref{eq:app_lorentz_components}) and
(\ref{eq:app_anis_components}) into Eq.~(\ref{eq:app_lin_vort}), and using
$\Delta p_0=\Delta\beta_0/2$, gives
\begin{align}
& \sigma \delta v_y =
ik B_0 \delta B_y+B_g'\delta B_z
-ikB_0B_g\delta\Delta p \nonumber\\
&\hspace{0.6in}
-\frac{\Delta\beta_0}{2}
\left[
ikB_0(1-2B_g^2)\delta B_y
+B_g'\delta B_z
+2B_0^2B_g\delta B_z'
\right]
+\nu\mathcal{L}\delta v_y,
\label{eq:app_dvy_final}\\
& \sigma\mathcal{L}\delta v_z =
ik\left(B_0\mathcal{L}\delta B_z-B_0''\delta B_z\right)
-k^2B_0\left(2B_0'\delta\Delta p+B_0\delta\Delta p'\right) \nonumber\\
&\hspace{0.6in}
-\frac{\Delta\beta_0}{2}
\Bigl\{
ik\left(B_0\mathcal{L}\delta B_z-B_0''\delta B_z\right)
-2\left[
ikB_0^2\left(B_0\delta B_z''+3B_0'\delta B_z'\right)\right. \nonumber\\
&\hspace{2.7in}
\left.
+k^2B_0^2B_g\delta B_y'
+k^2B_0\left(B_0B_g'+2B_0'B_g\right)\delta B_y
\right]
\Bigr\}
+\nu\mathcal{L}^2\delta v_z .
\label{eq:app_dvz_final}
\end{align}
These are Eqs.~(\ref{dvy}) and (\ref{dvz}).

\subsection{Linearized Pressure-Difference Equation}

The advective term in Eq.~(\ref{eq:app_incompressible_deltap}) does not
contribute at first order because $\mathbf{v}_0=0$ and
$\Delta p_0$ is uniform. The viscous heating term is also second order in the
perturbations because $\boldsymbol{\omega}_0=0$. The pressure-strain
contribution is
\begin{equation}
    \delta\left[
    \mathbf{b}\cdot(\mathbf{b}\cdot\nabla)\mathbf{v}
    \right]
    =
    B_0\left(ikB_g\delta v_y-B_0\delta v_z'\right),
\end{equation}
so the ideal pressure response is
\begin{align}
    \mathcal{P}_{\rm str}
    &=C_{p,0}B_0
    \left(B_0\delta v_z'-ikB_g\delta v_y\right), \\
    C_{p,0}
    &=
    \frac{1}{2}\left[
    (\gamma_\parallel+\gamma_\perp-2)\beta_0
    +(\gamma_\parallel-1)\Delta\beta_0
    \right].
\end{align}
For the ohmic part, define
\begin{equation}
    S_0\equiv\mathbf{B}_0\cdot\mathbf{J}_0
    =B_gB_0'-B_0B_g'.
\end{equation}
To first order,
\begin{equation}
    \delta(\mathbf{b}\cdot\mathbf{J})
    =
    \mathbf{B}_0\cdot\delta\mathbf{J}
    +\mathbf{J}_0\cdot\delta\mathbf{B}
    -S_0(\mathbf{B}_0\cdot\delta\mathbf{B}).
\end{equation}
After using Eq.~(\ref{eq:app_constraints}),
\begin{align}
    k\,\delta(\mathbf{b}\cdot\mathbf{J})
    &=
    k\left(B_0'\delta B_y-B_0\delta B_y'\right)
    +i\left[B_g\left(\delta B_z''-k^2\delta B_z\right)
    -B_g'\delta B_z'\right]
    -S_0\left(iB_0\delta B_z'+kB_g\delta B_y\right), \\
    k\,\mathbf{J}_0\cdot\delta\mathbf{J}
    &=
    kB_g'\delta B_y'
    +iB_0'\left(\delta B_z''-k^2\delta B_z\right).
\end{align}
Therefore the full linearized pressure-difference equation is
\begin{align}
& \sigma k\delta\Delta p =
C_{p,0}kB_0\left(B_0\delta v_z'-ikB_g\delta v_y\right) \nonumber\\
&\hspace{0.3in}
+2\eta G S_0
\left\{
k\left(B_0'\delta B_y-B_0\delta B_y'\right)
+i\left[B_g\left(\delta B_z''-k^2\delta B_z\right)-B_g'\delta B_z'\right]
-S_0\left(iB_0\delta B_z'+kB_g\delta B_y\right)
\right\} \nonumber\\
&\hspace{0.3in}
-2\eta H
\left[
kB_g'\delta B_y'
+iB_0'\left(\delta B_z''-k^2\delta B_z\right)
\right],
\label{eq:app_dp_final}
\end{align}
with $G=\gamma_\parallel+\gamma_\perp-2$ and
$H=\gamma_\perp-1$. Equation~(\ref{eq:app_dp_final}) is Eq.~(\ref{dp_lin})
with the two parts of $C_{p,0}$ written separately in the main text.

\subsection{Final Form for the Force-Free Harris Sheet}

Using the identities listed in Appendix~\ref{app:equilibrium-identities},
the derivative coefficients in
Eqs.~(\ref{eq:app_dvy_final})--(\ref{eq:app_dp_final}) can be eliminated.
The induction equations become
\begin{align}
    \sigma\delta B_y
    &=ikB_0\delta v_y+\frac{B_0B_g}{a}\delta v_z
    +\eta\mathcal{L}\delta B_y, \\
    \sigma\delta B_z
    &=ikB_0\delta v_z+\eta\mathcal{L}\delta B_z .
\end{align}
The velocity equations are
\begin{align}
& \sigma \delta v_y =
ik B_0 \delta B_y-\frac{B_0B_g}{a}\delta B_z
-ikB_0B_g\delta\Delta p \nonumber\\
&\hspace{0.6in}
-\frac{\Delta\beta_0}{2}
\left[
ikB_0(1-2B_g^2)\delta B_y
-\frac{B_0B_g}{a}\delta B_z
+2B_0^2B_g\delta B_z'
\right]
+\nu\mathcal{L}\delta v_y,
\label{eq:app_dvy_forcefree}\\
& \sigma\mathcal{L}\delta v_z =
ikB_0\left(\mathcal{L}\delta B_z+\frac{2B_g^2}{a^2}\delta B_z\right)
-k^2B_0\left(\frac{2B_g^2}{a}\delta\Delta p
+B_0\delta\Delta p'\right) \nonumber\\
&\hspace{0.6in}
-\frac{\Delta\beta_0}{2}kB_0
\Biggl\{
i\left[
\left(1-2B_0^2\right)\delta B_z''
-\frac{6B_0B_g^2}{a}\delta B_z'
+\left(\frac{2B_g^2}{a^2}-k^2\right)\delta B_z
\right]
\nonumber\\
&\hspace{1.05in}
-2kB_g\left[
B_0\delta B_y'
+\frac{1}{a}\left(2B_g^2-B_0^2\right)\delta B_y
\right]
\Biggr\}
+\nu\mathcal{L}^2\delta v_z .
\label{eq:app_dvz_forcefree}
\end{align}
The pressure-difference equation reduces to
\begin{align}
& \sigma k\delta\Delta p =
C_{p,0}kB_0\left(B_0\delta v_z'-ikB_g\delta v_y\right) \nonumber\\
&\hspace{0.3in}
+\frac{2\eta (\gamma_\parallel-1) B_g}{a}
\left[
-kB_0\delta B_y'
+iB_g\mathcal{L}\delta B_z
\right],
\label{eq:app_dp_forcefree}
\end{align}
Only the parallel pressure index remains in the resistive
pressure-difference response for this equilibrium. This is the same
linearized system as in the main text, now written with the force-free Harris
equilibrium and arbitrary current-sheet thickness $a$ substituted explicitly.

The same force-free system can be written with real coefficients only. The
factors of $i$ in Eqs.~(\ref{dvy})--(\ref{dp_lin}) are a consequence of the
Fourier convention $e^{ikx+\sigma t}$ and of the incompressibility and
solenoidal constraints in Eq.~(\ref{eq:app_constraints}). They can be removed
by the constant phase rotation
\begin{equation}
    \delta v_y=i u_y,\qquad
    \delta v_z=u_z,\qquad
    \delta B_y=b_y,\qquad
    \delta B_z=i b_z,\qquad
    \delta\Delta p=p_\Delta .
    \label{eq:real_variables}
\end{equation}
Equivalently,
$\delta\mathbf{q}=\mathsf{D}\mathbf{q}_{\rm r}$ with
$\mathsf{D}=\mathrm{diag}(i,1,1,i,1)$ and
$\mathbf{q}_{\rm r}=(u_y,u_z,b_y,b_z,p_\Delta)^T$. Since $\mathsf{D}$ is
nonsingular and independent of $z$, this transformation leaves the
generalized eigenvalues unchanged. It only fixes the relative quadrature
between the $x$-even and $x$-odd perturbation amplitudes. The eliminated
components have the corresponding phases
\begin{equation}
    \delta v_x=\frac{i}{k}u_z',
    \qquad
    \delta B_x=-\frac{1}{k}b_z' .
    \label{eq:real_eliminated_components}
\end{equation}
Using Eqs.~(\ref{eq:real_variables}) and
(\ref{eq:real_eliminated_components}), the force-free Harris-sheet system
becomes
\begin{align}
\sigma b_y
&=-kB_0u_y+\frac{B_0B_g}{a}u_z+\eta\mathcal{L}b_y,
\label{eq:real_forcefree_by}\\
\sigma b_z
&=kB_0u_z+\eta\mathcal{L}b_z,
\label{eq:real_forcefree_bz}\\
\sigma u_y
&=kB_0b_y-\frac{B_0B_g}{a}b_z-kB_0B_gp_\Delta \nonumber\\
&\quad
-\frac{\Delta\beta_0}{2}
\left[
kB_0(1-2B_g^2)b_y
-\frac{B_0B_g}{a}b_z
+2B_0^2B_gb_z'
\right]
+\nu\mathcal{L}u_y,
\label{eq:real_forcefree_uy}\\
\sigma\mathcal{L}u_z
&=-kB_0\left(\mathcal{L}b_z+\frac{2B_g^2}{a^2}b_z\right)
-k^2B_0\left(\frac{2B_g^2}{a}p_\Delta+B_0p_\Delta'\right) \nonumber\\
&\quad
+\frac{\Delta\beta_0}{2}kB_0
\Biggl\{
\left(1-2B_0^2\right)b_z''
-\frac{6B_0B_g^2}{a}b_z'
+\left(\frac{2B_g^2}{a^2}-k^2\right)b_z \nonumber\\
&\hspace{1.15in}
+2kB_g\left[
B_0b_y'
+\frac{1}{a}\left(2B_g^2-B_0^2\right)b_y
\right]
\Biggr\}
+\nu\mathcal{L}^2u_z,
\label{eq:real_forcefree_uz}\\
\sigma k p_\Delta
&=C_{p,0}kB_0\left(B_0u_z'+kB_g u_y\right) \nonumber\\
&\quad
-\frac{2\eta (\gamma_\parallel-1) B_g}{a}
\left[
kB_0b_y'
+B_g\mathcal{L}b_z
\right] .
\label{eq:real_forcefree_pi}
\end{align}
All coefficients in
Eqs.~(\ref{eq:real_forcefree_by})--(\ref{eq:real_forcefree_pi}) are real for
real equilibrium profiles, real $k$, and real transport coefficients.

\section{Double-Isothermal Limit of the Outer Problem}
\label{app:isothermal-outer}
\setcounter{equation}{0}

The double-isothermal closure,
$(\gamma_\parallel,\gamma_\perp)=(1,1)$, is special because the pressure
response coefficient in Eq.~(\ref{eq:barbeta}) vanishes:
\begin{equation}
    \bar{\beta}=0 .
\end{equation}
Equivalently, the ideal pressure-anisotropy perturbation in
Eq.~(\ref{eq:outer-explicit}) is not driven. In the force-free form of the
linearized equations, Eq.~(\ref{eq:app_dp_forcefree}) also has
$C_{p,0}=0$ and $\gamma_\parallel-1=0$, so the resistive source of
$\delta\Delta p$ vanishes as well. Thus the double-isothermal tearing problem
retains the prescribed equilibrium anisotropy but loses all explicit
dependence on the equilibrium plasma-$\beta$ value $\beta_0$.

The remaining anisotropy dependence is carried by
\begin{equation}
    \mathcal{A}=1-\frac{\Delta\beta_0}{2},
    \qquad
    \mathcal{R}_0=1+\frac{\Delta\beta_0}{2},
    \qquad
    q^2\equiv\frac{\mathcal{A}}{\mathcal{R}_0}.
    \label{eq:iso_parameters}
\end{equation}
The quasistatic localized branch requires $q^2>0$, or
$-2<\Delta\beta_0<2$, before imposing pressure positivity and additional
microinstability constraints. With $\zeta=z/a$, $\alpha=ka$, and
$b\equiv\delta B_z$, the double-isothermal outer equation can be written
using
\begin{equation}
    \mathcal{R}(\zeta)
    =
    \mathcal{A}+\Delta\beta_0\,\sech^2\zeta
    =
    \mathcal{R}_0-\Delta\beta_0\,\tanh^2\zeta .
    \label{eq:iso_R}
\end{equation}
Equation~(\ref{eq:outer-explicit}) then reduces to
\begin{align}
    0={}&b_{\zeta\zeta}
    +\frac{2\Delta\beta_0\tanh\zeta\,\sech^2\zeta}
    {\mathcal{R}}b_\zeta \nonumber\\
    &+\left[
    \frac{2\mathcal{A}\sech^2\zeta}{\mathcal{R}}
    -\alpha^2\frac{\mathcal{R}}{\mathcal{R}_0}
    \right]b .
    \label{eq:iso_outer_zeta}
\end{align}
This form is already independent of $\beta_0$. Since the first-derivative
coefficient is $-\mathcal{R}_\zeta/\mathcal{R}$,
Eq.~(\ref{eq:iso_outer_zeta}) is equivalently
\begin{equation}
    \frac{d}{d\zeta}
    \left(
    \frac{1}{\mathcal{R}}\frac{db}{d\zeta}
    \right)
    +\left[
    \frac{2\mathcal{A}\sech^2\zeta}{\mathcal{R}^2}
    -\frac{\alpha^2}{\mathcal{R}_0}
    \right]b=0 .
    \label{eq:iso_outer_self_adjoint}
\end{equation}

The far-field decay rate and the exact marginal point are
\begin{equation}
    \lambda a=\alpha q,
    \qquad
    \alpha_c=q.
    \label{eq:iso_alpha_c}
\end{equation}
At $\alpha=\alpha_c$, Eq.~(\ref{eq:iso_outer_zeta}) admits the exact even
solution
\begin{equation}
    b_m(\zeta)=\sech^{q^2}\zeta ,
    \label{eq:iso_marginal_solution}
\end{equation}
for which $b_m'(0)=0$ and therefore $\Delta'=0$. This is the
double-isothermal specialization of Eq.~(\ref{eq:exact_marginal_outer}).

The long-wavelength tearing index follows from the exact $\alpha=0$ solutions
\begin{equation}
    b_1=\tanh\zeta,
    \qquad
    b_2=\mathcal{A}\zeta\tanh\zeta-\mathcal{R}_0 .
    \label{eq:iso_zero_alpha_solutions}
\end{equation}
Matching their large-$\zeta$ behavior to
$e^{-\alpha q\zeta}\simeq1-\alpha q\zeta$ gives
\begin{equation}
    \Delta'
    \sim
    \frac{2\mathcal{A}}{(\lambda a)\mathcal{R}_0}
    =
    \frac{2q}{\alpha},
    \qquad
    \alpha\ll\alpha_c .
    \label{eq:iso_delta_long}
\end{equation}
Thus the leading long-wavelength coefficient is independent of $\beta_0$ and
depends on the equilibrium anisotropy only through the ratio
$q=(\mathcal{A}/\mathcal{R}_0)^{1/2}$.

A compact interpolation that preserves both the exact long-wavelength
coefficient and the exact marginal point is
\begin{equation}
    \Delta'
    \simeq
    2\left(\frac{q}{\alpha}-\frac{\alpha}{q}\right)
    =
    2\left[
    \frac{1}{\alpha}
    \sqrt{\frac{\mathcal{A}}{\mathcal{R}_0}}
    -
    \alpha
    \sqrt{\frac{\mathcal{R}_0}{\mathcal{A}}}
    \right].
    \label{eq:iso_delta_closure}
\end{equation}
This is the double-isothermal specialization of
Eq.~(\ref{eq:Delta_global_closure}) and reduces to the classical MHD result
when $\Delta\beta_0\to0$.

For finite wavelengths, the fitted curvature ratio $a_2/a_0$ introduced in
Appendix~\ref{app:outer-stability-boundary} gives the fitted tearing index
$\Delta'_{\rm fit}$ used in Figure~\ref{fig:dispersion-branches}. This fitted
outer input closely reproduces the numerical double-isothermal FKR branch.
The double-isothermal eigenvalue scans remain independent of $\beta_0$ because
their governing equations contain only $S$, $\alpha$, and the
equilibrium-anisotropy factors $\mathcal{A}$ and $\mathcal{R}_0$.

\section{Numerical Stability Boundary from the Outer Region}
\label{app:outer-stability-boundary}
\setcounter{equation}{0}

This appendix provides a direct numerical check of the marginal condition
derived in Eq.~(\ref{eq:alpha_marginal}). The calculation uses the reduced
ideal outer-region equation, Eq.~(\ref{eq:outer-explicit}), in the
dimensionless coordinate $\zeta=z/a$, with
$B_0=\tanh\zeta$, $B_g=\sech\zeta$, and $\alpha=ka$. For a specified
$(\beta_0,\Delta\beta_0)$ and closure
$(\gamma_\parallel,\gamma_\perp)$, the solution is integrated inward from an
asymptotic point $\zeta_{\max}=20$ using the decaying boundary condition
\begin{equation}
    b(\zeta_{\max})=\exp[-(\lambda a)\zeta_{\max}],
    \qquad
    b_\zeta(\zeta_{\max})
    =-(\lambda a)b(\zeta_{\max}),
\end{equation}
where
\begin{equation}
    \lambda a
    =
    \alpha\sqrt{\frac{\mathcal{A}}{\mathcal{R}_0}} .
\end{equation}
Here $b\equiv\delta B_z$ and $b_\zeta=db/d\zeta$. The integration therefore
uses only the far-field decay of the localized outer eigenfunction, not the
analytic marginal solution at the sheet center.

After the inward integration reaches $\zeta=0$, the numerical diagnostic is
\begin{equation}
    D(\alpha;\beta_0,\Delta\beta_0)
    \equiv
    \frac{b_\zeta(0)}{b(0)} .
\end{equation}
By reflection symmetry, $\Delta'=2D$. The marginal stability boundary is
therefore found by solving $D=0$ for $\alpha$, using a one-dimensional root
search in $\alpha$. In the numerical implementation, the inward integration is
performed with an adaptive DOP853 Runge--Kutta method with relative and
absolute tolerances $10^{-10}$ and $10^{-12}$, respectively, and the root is
obtained with a bracketing Brent method. Parameter values with
$\Delta\beta_0\leq-\beta_0$ are excluded because they would give
$p_{\parallel,0}\leq0$, and values for which $\mathcal{A}/\mathcal{R}_0<0$ are
excluded because the far-field decay rate is not real in the quasistatic outer
problem.

\begin{figure}[!ht]
\centering
\includegraphics[width=0.75\textwidth]{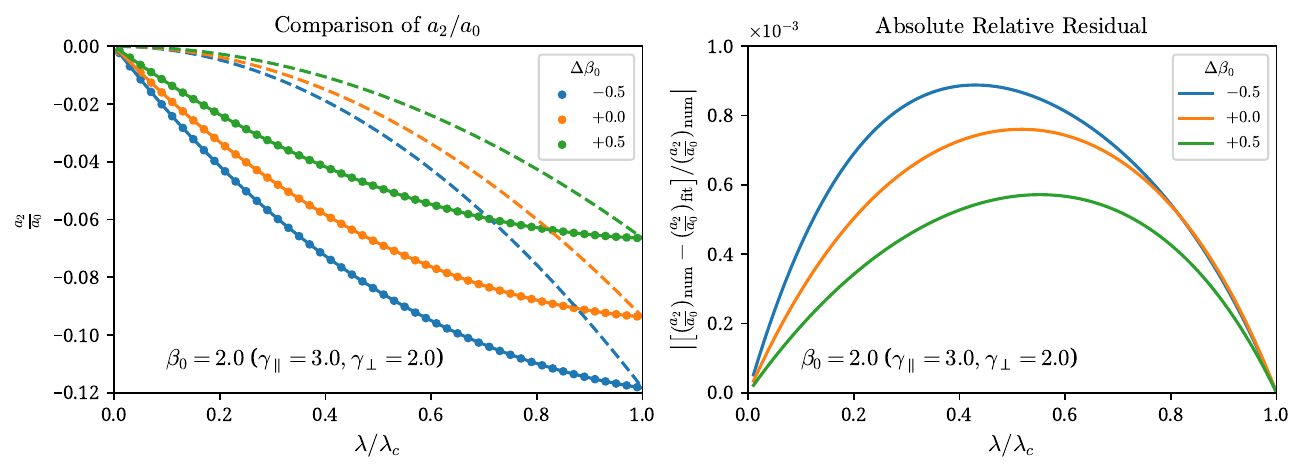}
\caption{Outer-region validation of the curvature coefficient
$a_2/a_0$ entering the tearing index. The left panel compares the directly
integrated outer solution (symbols), the compact theoretical estimate from
Eq.~(\ref{eq:a2a0_global_closure}) (dashed curves), and the empirical
interpolation from Eq.~(\ref{eq:a2a0_empirical_fit}) (solid curves), for the
double-adiabatic closure at $\beta_0=2$ and
$\Delta\beta_0=-0.5$, $0$, and $+0.5$. The right panel shows the absolute
relative residual between the numerical and fitted values; the plotted scale is
$10^{-3}$.}
\label{fig:a2a0-fit-comparison}
\end{figure}

For future comparison with the local series solution, we also record a
Pad\'e-like empirical approximation for the global value of the curvature
coefficient $a_2/a_0$. Let
\begin{equation}
    \ell\equiv\lambda a
    =
    \alpha\sqrt{\frac{\mathcal{A}}{\mathcal{R}_0}} .
\end{equation}
Then the generalized fit is
\begin{equation}
    \left(\frac{a_2}{a_0}\right)_{\rm fit}
    =
    \mu\ell
    \left[
    \frac{1}{2}
    \frac{\ell\left(\mathcal{R}_0^{-1}+c/\mathcal{A}\right)}
    {\ell+c}
    -\mathcal{R}_0^{-1}
    \right],
    \qquad
    c\simeq\frac{5}{4}.
    \label{eq:a2a0_empirical_fit}
\end{equation}
The Pad\'e-like structure in Eq.~(\ref{eq:a2a0_empirical_fit}) was selected
using the SymbolicRegression.jl symbolic-regression package
\citep{Cranmer2023SymbolicRegression}, applied to numerical outer-region
integrations spanning several values of $\beta_0$ and $\Delta\beta_0$. The
value $c=5/4$ was then adopted as a single heuristic calibration from these
integrations; it was not re-fitted separately for each dispersion curve,
closure, or anisotropy value. Its role is to improve the finite-wavelength
interpolation, while the long-wavelength coefficient and the marginal value
are enforced by the analytic constraints described above.
Because a single calibration is used beyond the subset displayed below,
$\Delta'_{\rm fit}$ should be treated as a heuristic finite-wavelength
interpolation rather than a uniformly validated approximation over the full
parameter domain.
For $\Delta\beta_0=0$, where $\mathcal{A}=1$,
$\mu=\bar{\beta}$, and $\mathcal{R}_0=1+\bar{\beta}$, this reduces to the
one-parameter form used to fit the isotropic-equilibrium scans. The factor
$c/\mathcal{A}$ is chosen so that the fit keeps the exact marginal value from
Eq.~(\ref{eq:a2a0_marginal_exact}). Indeed, at
$\alpha=\alpha_c$, one has $\ell_c=\mathcal{A}/\mathcal{R}_0$, and
Eq.~(\ref{eq:a2a0_empirical_fit}) gives
\begin{equation}
    \left.
    \left(\frac{a_2}{a_0}\right)_{\rm fit}
    \right|_{\alpha=\alpha_c}
    =
    -\frac{\mathcal{A}\mu}{2\mathcal{R}_0^2}.
    \label{eq:a2a0_empirical_marginal}
\end{equation}
Substitution of Eq.~(\ref{eq:a2a0_empirical_fit}) into
Eq.~(\ref{eq:Delta_full}) gives the corresponding fitted tearing index,
\begin{align}
    \Delta'_{\rm fit}
    &=
    \frac{2\mathcal{A}}{\mathcal{R}_0\ell}
    -\frac{2\mu}{\mathcal{R}_0}
    +\frac{\mu\ell}{\ell+c}
    \left(
    \mathcal{R}_0^{-1}
    +\frac{c}{\mathcal{A}}
    \right)
    -\left(2+\frac{\mu}{\mathcal{A}}\right)\ell .
    \label{eq:Delta_empirical_fit}
\end{align}
Equivalently, written directly in terms of $\alpha$,
\begin{align}
    \Delta'_{\rm fit}
    &=
    \frac{2}{\alpha}
    \sqrt{\frac{\mathcal{A}}{\mathcal{R}_0}}
    -\frac{2\mu}{\mathcal{R}_0}
    +\frac{
    \mu\alpha\sqrt{\mathcal{A}/\mathcal{R}_0}
    }{
    c+\alpha\sqrt{\mathcal{A}/\mathcal{R}_0}
    }
    \left(
    \mathcal{R}_0^{-1}
    +\frac{c}{\mathcal{A}}
    \right) \nonumber\\
    &\quad
    -\left(2+\frac{\mu}{\mathcal{A}}\right)
    \alpha\sqrt{\frac{\mathcal{A}}{\mathcal{R}_0}} .
    \label{eq:Delta_empirical_fit_alpha}
\end{align}
Because Eq.~(\ref{eq:a2a0_empirical_fit}) preserves the exact marginal value
of $a_2/a_0$, Eq.~(\ref{eq:Delta_empirical_fit}) also satisfies
$\Delta'_{\rm fit}=0$ at
$\alpha=\alpha_c=\sqrt{\mathcal{A}/\mathcal{R}_0}$.
It also satisfies $a_2/a_0\to0$ as $\ell\to0$, but with a leading linear
dependence on $\ell$ rather than the parabolic behavior assumed in
Eq.~(\ref{eq:a2a0_global_closure}). Equation~(\ref{eq:a2a0_empirical_fit})
should therefore be regarded as a testable global interpolation for the
decaying outer solution, not as a result of a finite polynomial truncation of
the local series.

\begin{figure}[!ht]
\centering
\includegraphics[width=0.32\textwidth]{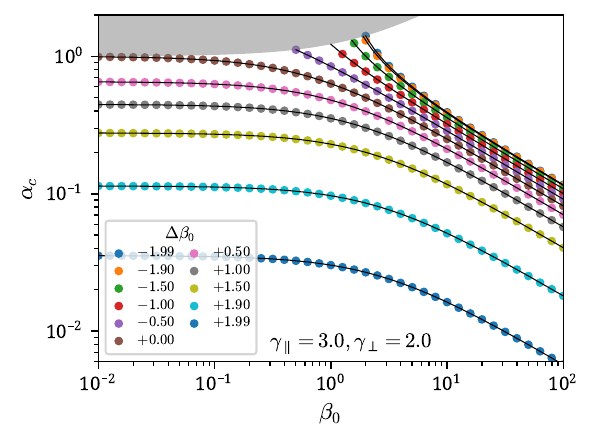}
\hfill
\includegraphics[width=0.32\textwidth]{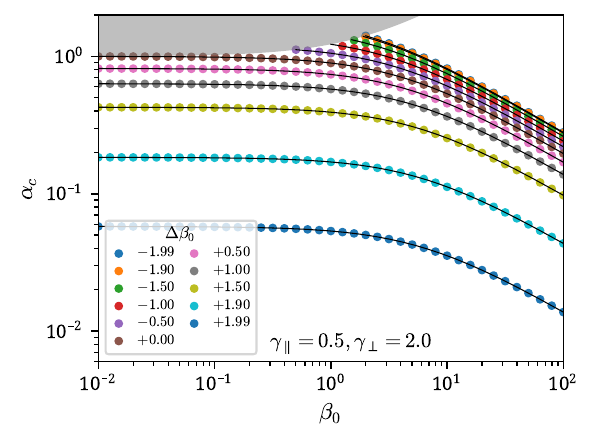}
\hfill
\includegraphics[width=0.32\textwidth]{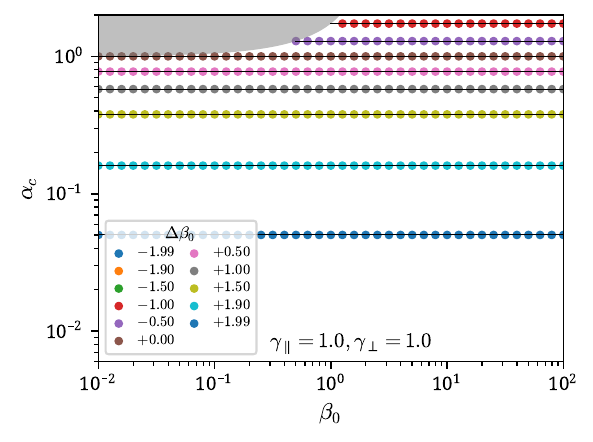}
\caption{Numerical outer-region stability boundary
$\alpha_c=k_c a$ as a function of $\beta_0$ for different imposed
anisotropies $\Delta\beta_0$. The left, center, and right panels correspond to
$(\gamma_\parallel,\gamma_\perp)=(3,2)$, $(0.5,2)$, and $(1,1)$,
respectively. Symbols show the roots of
$b_\zeta(0)/b(0)=0$ obtained by inward integration of the outer equation.
The solid black curves show the analytic boundary
$\alpha_c=\sqrt{\mathcal{A}/\mathcal{R}_0}$ from
Eq.~(\ref{eq:alpha_marginal}). The grey shaded region is outside the
pressure-positive domain mapped through the same analytic boundary.}
\label{fig:outer-stability-boundary}
\end{figure}

Figure~\ref{fig:a2a0-fit-comparison} compares this fitted interpolation
with a direct inward integration of the outer equation. The comparison is made
for the double-adiabatic closure at $\beta_0=2$ and
$\Delta\beta_0=-0.5$, $0$, and $+0.5$. We plot the result as a function of
$\ell/\ell_c$, where $\ell=\lambda a$ and
$\ell_c=\mathcal{A}/\mathcal{R}_0$ is the marginal value corresponding to
$\alpha=\alpha_c$. Since Eq.~(\ref{eq:Delta_full}) maps $a_2/a_0$
algebraically into $\Delta'$, this comparison tests the nontrivial
outer-region input used to construct $\Delta'_{\rm fit}$. The simple
theoretical estimate in Eq.~(\ref{eq:a2a0_global_closure}) preserves the
long-wavelength and marginal constraints but underestimates the finite
wavelength curvature, whereas the empirical interpolation in
Eq.~(\ref{eq:a2a0_empirical_fit}) tracks the numerical outer solution with a
relative residual below $10^{-3}$ over the plotted range. This displayed test
does not establish the same residual bound for the other closures or for the
full anisotropy range used in the dispersion scans; in those cases the fitted
FKR curves provide a consistency comparison rather than a separately validated
outer solution.

Figure~\ref{fig:outer-stability-boundary} shows that the numerically obtained
roots of $D=0$ lie on the analytic curves
$\alpha_c=\sqrt{\mathcal{A}/\mathcal{R}_0}$ for all three closures and over
the scanned values of $\beta_0$ and $\Delta\beta_0$. This agreement confirms
that the true marginal boundary is set by Eq.~(\ref{eq:alpha_marginal}). It
also verifies that the marginal point is controlled by the global decaying
outer solution, rather than by a purely local truncation of the series near
the resonant surface. Thus the tearing-unstable localized branch satisfies
$\alpha<\alpha_c$, with $\alpha_c$ given by
Eq.~(\ref{eq:alpha_marginal}).

\bibliography{ms}{}
\bibliographystyle{aasjournal}

\end{document}